\documentclass[11pt,a4paper]{article}

\usepackage[latin1]{inputenc}
\usepackage{tikz}
\usetikzlibrary{trees}

\usepackage[tbtags]{amsmath}    
\usepackage{amsfonts,amssymb,amsthm,euscript,makeidx,pifont,boxedminipage,multicol,fullpage}
\usepackage{color}
\usepackage{yfonts}

\usepackage{comment,todonotes}

\usepackage[latin1]{inputenc}
\usepackage{tikz}
\usetikzlibrary{trees}

\usepackage{bbm}
\usepackage{t1enc}
\usepackage{amsfonts,amssymb}
\usepackage{latexsym,amsmath,amscd}
\usepackage{enumerate}
\usepackage{color}
\usepackage{hyperref}
\usepackage{mathrsfs}
\usepackage{amsthm}

\newcommand{\be}{\begin{equation}}
\newcommand{\ee}{\end{equation}}
\newcommand{\bde}{\begin{displaymath}}
\newcommand{\ede}{\end{displaymath}}
\newcommand{\beq}{\begin{eqnarray*}}
\newcommand{\eeq}{\end{eqnarray*}}
\newcommand{\beqa}{\begin{eqnarray}}
\newcommand{\eeqa}{\end{eqnarray}}

\newtheorem{Theorem}{Theorem}[section]
\newtheorem{Proposition}[Theorem]{Proposition}
\newtheorem{ass}[Theorem]{Assumption}
\newtheorem{sass}[Theorem]{Standing Assumption}
\newtheorem {Cor}[Theorem]{Corollary}
\newtheorem {pro}[Theorem]{Proposition}

\newtheorem {Lemma}[Theorem]{Lemma}
\newtheorem {rem}[Theorem]{Remark}
\newtheorem {rems}[Theorem]{Remarks}
\newtheorem {com}[Theorem]{Comment}
\newtheorem {coms}[Theorem]{Comments}
\newtheorem {Definition}[Theorem]{Definition}
\theoremstyle{definition}
\newtheorem {exam}[Theorem]{Example}

\newcommand{\bnota}{\begin{notation} \rm } \newcommand{\enota}{\end{notation}}
\newcommand{\bcom}{\begin{com} \rm } \newcommand{\ecom}{\end{com}}
\newcommand{\bcoms}{\begin{coms} \rm } \newcommand{\ecoms}{\end{coms}}
\newcommand {\bdefi}{\begin{Definition}}
\newcommand {\edefi}{\end{Definition}}
\newcommand {\bl}{\begin{Lemma}}
\newcommand {\el}{\end{Lemma}}
\newcommand {\bethe}{\begin{Theorem}}
\newcommand {\eethe}{\end{Theorem}}
\newcommand {\bp}{\begin{pro}}
\newcommand {\ep}{\end{pro}}
\newcommand {\bcor}{\begin{Cor}}
\newcommand {\ecor}{\end{Cor}}
 \newcommand {\brem }{\begin{rem} \rm }
\newcommand {\erem }{\end{rem}}
 \newcommand {\brems }{\begin{rems} \rm }
\newcommand {\erems }{\end{rems}}

 \def\d{D}
 
  \def\F{{\mathcal F}}
\def\G{{\mathcal G}}
\def\H{{\mathcal H}}

\def\I{{\Omega}}
\def\timech{\kappa}
\def\ff{{\mathbb F}}
\def\gg{{\mathbb G}}
\def\hh{{\mathbb H}}

\def\gg{{\mathbb G}}
\def\hh{{\mathbb H}}
\def\Q{{\mathbb Q}}
\def\P{{\mathbb P}}
\def\R{{\mathbb R}}
\def\E {\mathbb{E}}

\def\No{\Lambda}
\def\no{\lambda}
\def\mo{\alpha}
\def\bet{\Lambda_0}
\def\Mo{\mathcal U}

\def\ale{\mathfrak a}

\def\norm#1{|| #1||}

\def \id{1\!\!1}

\newcommand{\is}{\circ}

\def \wh  {\widehat }
\def \wt  {\widetilde }
\def \proof {{\sc{Proof:}}~}
\def \finproof {\hfill $\square$  \\ }

\newcommand{\vf}[2][\I]{V^\ff_{\mathcal X, \mathcal P, #1}(#2)}
\newcommand{\tvf}[2][\I]{\wt V^\ff_{\mathcal X, \mathcal P, #1}(#2)}
\newcommand{\vg}[2][\I]{V^{\gg^-}_{\mathcal X, \mathcal P, #1}(#2)}
\newcommand{\vgg}[2][\I]{V^{\gg^+}_{\mathcal X, \mathcal P, #1}(#2)}
\newcommand{\vggg}[2][\I]{V^{\gg}_{\mathcal X, \mathcal P, #1}(#2)}

\newcommand{\pf}[2][\I]{P^\ff_{\mathcal X, \mathcal P, #1}(#2)}
\newcommand{\tpf}[2][\I]{\wt P^\ff_{\mathcal X, \mathcal P, #1}(#2)}

\newcommand{\pg}[2][\I]{P^{\gg^-}_{\mathcal X, \mathcal P, #1}(#2)}
\newcommand{\pgg}[2][\I]{P^{\gg^+}_{\mathcal X, \mathcal P, #1}(#2)}
\newcommand{\pggg}[2][\I]{P^{\gg}_{\mathcal X, \mathcal P, #1}(#2)}

\newcommand{\indicators}[1]{\mathbbm{1}_{{#1}}}

\newcommand{\Xc}{\mathcal{X}}
\newcommand{\Pc}{\mathcal{P}}
\newcommand{\Ac}{\mathcal{A}}
\newcommand{\Mcal}{\mathcal{M}}

\begin{document}

\title
{Robust framework for quantifying the value of information in pricing and hedging\thanks{The project has been generously supported by the European Research Council under the European Union's Seventh Framework Programme (FP7/2007-2013) / ERC grant agreement no. 335421. The authors also acknowledge the support of the Oxford-Man Institute of Quantitative Finance. 
Zhaoxu Hou is further grateful to Balliol College in Oxford and Jan Ob\l\'oj to St John's College in Oxford for their financial support.}}



\author{Anna Aksamit\thanks{E-mail: anna.aksamit@maths.ox.ac.uk}, Zhaoxu Hou\thanks{E-mail: zhaoxu.hou@maths.ox.ac.uk} \, and Jan Ob\l\'{o}j\thanks{E-mail: jan.obloj@maths.ox.ac.uk}}

\date{\today}

\maketitle

\begin{abstract}
We investigate asymmetry of information in the context of robust approach to pricing and hedging of financial derivatives. We consider two agents, one who only observes the stock prices and another with some additional information, and investigate when the pricing--hedging duality for the former extends to the latter. We introduce a general framework to express the superhedging and market model prices for an informed agent. Our key insight is that an informed agent can be seen as a regular agent who can restrict her attention to a certain subset of possible paths. 
We use results of Hou \& Ob\l\'oj \cite{ho_beliefs} on robust approach with beliefs to establish the pricing--hedging duality for an informed agent. Our results cover number of scenarios, including information arriving before trading starts, arriving after static position in European options is formed but before dynamic trading starts or arriving at some point before the maturity. For the latter we show that the superhedging value satisfies a suitable dynamic programming principle, which is of independent interest.
\vspace{0.3cm}


\end{abstract}

\section{Introduction}

Robust approach to pricing and hedging has been an active field of research in mathematical finance over the recent years. 
In this approach, instead of choosing a single probabilistic model, one considers superhedging simultaneously under a family of models, or pathwise on a set of feasible trajectories.  
Typically dynamic trading strategy in stocks and static trading, i.e.\ at time zero, in some European options are allowed. This setup was pioneered in the seminal work of Hobson \cite{Hobson} who obtained robust pricing and hedging bounds for lookback options. Therein, all martingale models which calibrate to the given market option prices, and superhedging for all canonical paths, were considered. Similar approach and setup, both in continuous and discrete time, were used to study other derivatives and abstract pricing--hedging duality questions, see e.g.\ 
\cite{BHR:01,cox2011robust,CoxObloj:11,CoxWang:11,acciaio2013model,mass_transport, bbkn:14,Yan,dolinsky2014martingale,bfm:15} and the references therein. Other papers, often intertwined with the previous stream, focused on the case when superhedging is only required under a given family of probability measures or, more recently, on a given set of feasible paths, see e.g.\ \cite{Lyons:95,AvellanedaLevyParas:95,Mykland:03,DenisMartini:06,bn:15,ho_beliefs} and the references therein.

The prevailing focus in the literature has been so far on the case when trading strategies are adapted to the natural filtration $\ff$ of the price process $S$. In contrast, our main interest in this paper is to understand what happens when a larger filtration $\gg$ is considered. One can think of $\ff$ as the filtration of an unsophisticated or small agent and $\gg$ as the filtration of a sophisticated agent who invests in acquiring additional information. Equally, $\gg$ may correspond to the filtration of an insider. In general, the two filtrations model an asymmetry of information between the agents. We are primarily interested in the case when the additional information does not lead to instant arbitrage opportunities but does offer an advantage and we wish to quantify this advantage in the context of robust pricing and hedging of derivatives. 

We develop pathwise approach. Our key insight is that the pricing and hedging problem for the agent with information in $\gg$ can often be reduced to that for the standard agent who only considers a subset of the pathspace\footnote{When working on this paper we were made aware of a related forthcoming paper \cite{ACHinfo_paper} which also considers informed agents using pathwise restrictions. However the technical setup therein is closely related to the pathwise approach developed in \cite{BCHPP}, based on Vovk's outer measure, and the paper develops a monotonicity principle in a similar sprit to \cite{BeiglbockCoxHuesmann}.}. 
This allows us to use the duality results with beliefs obtained by Hou and Ob\l\'oj \cite{ho_beliefs}. 
Specifically, we consider the price process as the canonical process on a restriction of the space of $\R^d$-valued continuous functions on $[0, T]$. 
Price process represents assets, stocks or options, which are traded continuously, subject to usual admissibility constraints, see Definition \ref{def:admissibility} below. We further allow static positions in a given set of options $\mathcal X$ whose market prices $\mathcal P$ are known. These are less liquid options which are not assumed to be traded after time zero. The additional information could arrive both before and/or after the static trading is executed. To account for this we add to $\gg$ an additional element $\G_{-1}$, $\{\emptyset, \I\}\subset \G_{-1}\subset \G_0$ and require that the static position $\alpha$ is $\G_{-1}$-measurable. The initial cost of such a position is $\mo \mathcal{P}(\Xc)$ and hence the superhedging cost of a derivative with payoff $\xi$, for an agent with filtration $\gg$, is given by
\begin{align*}
\vggg[\I]{\xi}(\omega):=\inf\Big\{\mo(\omega)\mathcal P(\Xc):&\, \, \exists \,\textrm{ $\gg$-admissible }\,(\mo, \gamma) \, \textrm{ s.t. }\, \mo\Xc+\int_0^T\gamma_u dS_u\geq \xi \,\, \textrm{on } \Omega\Big\},
\end{align*} 
where $\alpha$ also includes a position in cash. The pricing counterpart is obtained via
$$
\pggg[\I]{\xi}(\omega):=\sup_{\P\in \mathcal M^\gg_{\Xc,\Pc,\I}}\E_{\P}[\xi|\G_{-1}](\omega),\quad
$$
where the supremum is taken over $\gg$-martingale measures calibrated to the market prices $\Pc$ of options $\Xc$ and where we take a suitable version of the conditional expectations. We show that both $\vggg[\I]{\xi}$ and $\pggg[\I]{\xi}$ are well defined and constant on atoms of $\G_{-1}$. 

We first focus on the case of an initial enlargement: $\G_0 = \sigma(Z)$, for some $\F_T$-measurable random variable $Z$. Initially, we treat the case when $\G_{-1}=\G_0$ and show that the duality for the informed agent is the same as duality for uninformed agents in Hou \& Ob\l\'oj \cite{ho_beliefs} with beliefs of the form $\{\omega: Z(\omega)=c\}$. 
We discuss two examples: a very specific information $Z=\sup_{t\in [0,T]}|S_t-1|$ and a rather vague information $Z=\id_{\{S_t\in(a,b) \; \forall t\in [0,T]\}}$. Subsequently, we show that the duality also extends to the case of trivial $\G_{-1}$, under mild technical assumptions on $Z$. 

Secondly, we focus on the case when  the additional information is disclosed at some time $T_1\in (0,T)$, i.e.\ the filtration $\gg$ is of the form: $\G_t=\F_t$ for $t\in[0,T_1)$ and $\G_t=\F_t\lor \sigma(Z)$ for $t\in [T_1,T]$. 
To prove the pricing--hedging duality we establish a dynamic programming principle for both the superhedging cost $V(\xi)$ and the market model price $P(\xi)$. These are of independent interest even in the case of $\ff=\gg$. 

We note that the pricing problem for an informed agent was also examined recently in \cite{acciaio2015semi}. 
However the focus therein is very different to ours. The authors do not study the pricing--hedging duality but instead focus on the pricing aspect $\pggg{\xi}$ with a trivial $\G_{-1}$. They show that it is enough to optimise over extreme measures and characterise these, in analogy to the seminal work of Jacod-Yor \cite{jy77}, as the ones under which semi--static completeness holds, i.e.\ perfect hedging of all suitably integrable $\xi$ using the underlying assets and statically traded derivatives in $\Xc$. Under some further assumptions on $\gg$, semi--static completeness under $\Q$ is equivalent to filtrations $\ff$ and $\gg$ coinciding under $\Q$. 

Our paper is organised as follows. In Section \ref{s:general} we define our robust pricing and hedging setup. 
In Section \ref{s:ph} we define the relevant pricing and hedging notions for the informed agents and establish their characterisations via pathspace restrictions.  
In Section \ref{s:duality} we present our main results on pricing--hedging duality under an initially enlarged filtration.  Theorem \ref{duality} treats the case of instantly available information and Theorem \ref{gminus} the case when the information may not be used to construct static portfolios. In Section \ref{s:dpp} we study the dynamic situation when the additional information arrives at some time $T_1\in (0,T)$. We establish relevant dynamic programming principles in Propositions \ref{dpV} and \ref{dpP}, and then we obtain pricing--hedging duality result in Theorem \ref{dp}. Finally, in Section \ref{sec:infovalue}, we propose a way to value the additional information and discuss how agent's valuation changes with the timing of information's arrival.

\section{General set-up}
\label{s:general}

\subsection{Traded assets}

We consider a financial market with $d+1$ underlying assets: a numeraire and $d\in \mathbb N$ risky underlying assets. 
We work in a frictionless setting with no transaction costs. 
The prices are denominated in the units of a numeraire, e.g.\ bank account, whose price is thus normalised to 1. 
We suppose that the prices of the $d$ risky underlying assets belong to the space $\Omega^d:=C_1([0,T], \R^d_+)$, i.e., the space of non-negative continuous functions $f$ from $[0, T]$ into $\R^d_+$ such that $f_0=(1,...,1)$. We endow $\Omega^d$ with the sup norm $\norm{f}_d:=\sup_{t\leq T}|f_t|$ with $|f_t|:=\sup_{1\leq n \leq d}|f^n_t|$ so that $(\Omega^d, \norm{\cdot}_d)$ is a Polish space. We will denote by $\F^d_T$ the Borel $\sigma$-field of subsets of $(\Omega^d, \norm{\cdot}_d)$ which is the $\sigma$-field generated by canonical process on $\Omega^d$. 

Apart from the risky underlying assets, there may be derivative products traded on the market. Some of these could be particularly liquid -- we will assume they trade dynamically -- and others may be less liquid and only available for trading at the initial time. We only consider European derivatives which can be seen as $\F^d_T$--measurable functions $X: \Omega^d\to \mathbb R$. 
In line with \cite{ho_beliefs}, we further assume all the payoffs $X$ are bounded and uniformly continuous. The options traded are assumed to have a well defined price at time zero given by a linear operator $\Pc$, i.e., $\Pc(X)$ is the price at time zero of the $\F^d_T$-measurable, bounded and uniformly continuous payoff $X$. 
We let $\Xc=(\Xc^\no)_{\no\in \No}$, where $\No$ is a set of an arbitrary cardinality, be the vector of the payoffs of market options available for static trading. We assume that $0\in\No$ and $\Xc^0$ is a unit of the numeraire, $\Xc^0(\omega)=1$ for each $\omega\in\I$, so $\Pc(\Xc^0)=1$.
There are further $K$ options, $K\geq 0$, with non-negative payoffs $X^{(c)}_1,..., X^{(c)}_K$, which are traded dynamically. We normalise their payoffs so that their initial prices are equal to 1 and we model this situation by augmenting the set of risky assets. We simply consider $d+K$ assets that may be traded at any time, $d$ underlyings and $K$ options, with paths in an extended path space, $\Omega^{d+K}$, with the norm $\norm{\cdot}:=\norm{\cdot}_{d+K}$. Specifically, as the prices at maturity $T$ must be consistent with payoffs, only the following subset of $\Omega^{d+K}$ is considered:
$$
\I:=\left\{\omega\in \Omega^{d+K} : \omega^{(d+i)}_T=X^{(c)}_i(\omega^{(1)},...,\omega^{(d)})/ \mathcal P(X^{(c)}_i)\quad \forall i\leq K\right\}.
$$
The space $\I$ was called the \emph{information space} in \cite{ho_beliefs} since it encodes the information about the initial prices and the payoffs of continuously traded options.
We note that $(\I, \norm{\cdot})$ is a Polish space since it is a closed subset of the Polish space $(\Omega^{d+K}, \norm{\cdot})$.

\subsection{Information}

Let $S$ be the canonical process on $\I$, i.e., $S_t(\omega):=\omega_t$.
We introduce the filtration $\ff:=(\F_t)_{t\leq T}$ generated by $S$, i.e., 
$\F_t:=\sigma(S_s: s\leq t)$ for each $t\in[0, T]$. 
Note that it is not a right--continuous filtration  and each element $\F_t$, or more generally $\F_\tau$ for a stopping time $\tau$, is countably generated as the Borel $\sigma$-field on a Polish space. 

We also consider an enlarged filtration $\gg:=(\G_t)_{t\leq T}$ defined as $\G_t:=\F_t\lor \H_t$, for each $t\in[0,T]$, where $\hh:=(\H_t)_{t\leq T}$ is another filtration. 
In the literature, e.g.\ \cite{j, jy:gf, mansuyyor}, typically two special cases of the filtration $\hh$ are studied. 
First, the filtration $\hh$ can be taken constant, $\H_t=\sigma(Z)$ where $Z$ is a random variable, and $\gg$ is then called the initial enlargement of $\ff$ with $Z$. 
Second, the filtration $\hh$ can be taken as $\H_t=\sigma(\id_{\{\tau\leq s\}}: s\leq t)$ where $\tau$ is a random time (a non-negative random variable) and $\gg$ is then called the progressive enlargement of $\ff$ with a random time $\tau$. 

In our considerations it is important to specify \emph{when} the additional information arrives -- some decisions may have to be taken before and some after the additional information is acquired. To model such situations, we add to each filtration $\gg$ an additional element $\G_{-1}$ with $\{\emptyset, \I\}\subset \G_{-1}\subset \G_0$. 
For a given arbitrary filtration $\gg$ we denote by $\gg^+$ the filtration such that $\G^+_{-1}=\G_0$ and $\G^+_t=\G_t$ for $t\in [0,T]$. Similarly, we denote by $\gg^-$ the filtration such that $\G^-_{-1}=\{\emptyset, \I\}$ and $\G^-_t=\G_t$ for $t\in [0,T]$. We note that for the natural filtration of the price process $\ff$ the only choice is $\F_{-1}=\F_0=\{\emptyset, \I\}$. Finally, we make the following
\begin{sass}
\label{as:1}
All $\sigma$-fields $\G_t$, $t\in\{-1\}\cup[0,T]$, in the enlarged filtration $\gg$ are countably generated.
\end{sass}

\subsection{Trading strategies}

We now discuss the notion of atoms of a $\sigma$-field and introduce the right class of trading strategies with respect to a  general filtration $\gg$. We refer to Dellacherie \& Meyer \cite[Ch.1, $\S$9-12]{dell14} for useful details.
For a measurable space $(\I, \F_T)$ and a sub $\sigma$-field $\G\subset \F_T$ we introduce the following equivalence relation.
\bdefi
\label{atom}
Let $\omega$ and $\wt \omega$ be two elements of $\I$, and $\G\subset \F_T$ be a $\sigma$-field. Then we say that $\omega$ and $\wt \omega$ are $\G$-equivalent, and write $\omega \sim_{\G} \wt \omega$, if for each $G\in \G$ we have $\id_G(\omega)=\id_G(\wt \omega)$.
\edefi

We call $\G$-atoms the equivalence classes in $\I$ with respect to this relation.
We denote by $A^\omega$ the atom which contains $\omega$: $A^\omega=\bigcap\{A: {A\in \G, \omega\in A}\}$.
Note that if $\G$ is countably generated, $\G=\sigma(B_n:n\geq 1)$, then each atom is an element of $\G$ as $A^\omega=\bigcap_n C_n$ is a countable intersection, where $C_n=B_n$ if $\omega\in B_n$ and $C_n=B^c_n$ if $\omega\in B^c_n$. Also, it is then enough to check the relation from Definition \ref{atom} on the generators of $\G$. Finally, we note that in our setting, $\omega \sim_{\F_t} \wt \omega$ if and only if $\omega_u=\wt \omega_u$ for each $u\leq t$; and $\omega \sim_{\sigma(Z)} \wt \omega$ if and only if $Z(\omega)=Z(\wt \omega)$.

In Definitions \ref{def:admissibility} and \ref{def:semitrading} we shall define trading in such a way that strategies can be chosen separately on each atom of $\G_0$ for dynamic trading and each atom of $\G_{-1}$ for static trading. 
In particular we do not require that the strategies are measurable on the entire $\I$ but allow for trading strategies which are measurable when restricted to appropriate atoms. It is expressed via the set of mappings 
\begin{equation}
\label{eq:UG}
\Mo(\G):=\{M:\I\to\R \textrm{ s.t. } \omega\sim_{\G}\wt \omega \textrm{ implies } M(\omega)=M(\wt \omega)\}
\end{equation}
where $\G\in\{\G_0, \G_{-1}\}$, which are only required to be constant on atoms of $\G$. In particular, using information contained in $\G$ does not enforce measurability. Information provided by the $\sigma$-field $\G$ might be seen as a signal or equivalently as a partition of $\I$ into atoms. 
After receiving a signal, an agent may decide to use it according to $\Mo(\G)$ which is usually a much larger set of actions than the $\G$-measurable mappings. We believe the above is an important point and our approach is inspired by \cite{DE04} and \cite{HBKM13} where the set of informed sets is distinguished from a $\sigma$-field, the former allowing for uncountable sums.

We only consider trading strategies $\gamma$ which are of finite variation. This allows, similarly to \cite{Yan,ho_beliefs}, to define integrals pathwise simply via the integration by parts formula:
$$\int_0^t\gamma(u)d\omega(u):=\gamma(t)\omega(t)-\gamma(0)\omega(0)-\int_0^t\omega(u)d\gamma(u),\quad \omega\in \I.$$
The above integration provides dynamic trading in continuously traded assets as defined below.
\bdefi
\label{def:admissibility}
(i) The mapping $\gamma: \I \times [0,T]\to \R^{d+K}$ is called
\begin{enumerate}
\item[(a)] c\`adl\`ag if for each $\omega\in \I$, $\gamma(\omega):[0,T]\to \R^{d+K}$ is c\`adl\`ag,
\item[(b)] $\gg$-adapted on atoms of $\G_0$ if for each atom $A^\omega$ of $\G_0$ the mapping $\gamma \id_{A^\omega}$ is $\gg$-adapted.
\end{enumerate}
(ii) 
The mapping $\gamma: \I \times [0,T]\to \R^{d+K}$ is called $(\gg,M)$-admissible for $M\in\Mo(\G_0)$, where $\Mo(\G_0)$ is defined in \eqref{eq:UG}, if it is c\`adl\`ag, $\gg$-adapted on atoms of $\G_0$ and of finite variation, satisfying 
\be\label{eq:admissibility}
\int_0^t\gamma(\omega)_u dS_u(\omega)\geq -M(\omega) \quad \forall\;t\in [0,T],\; \omega\in \I\, .
\ee 
The set of all $(\gg,M)$-admissible strategies is denoted $\Ac^M(\gg)$. The set of all $\gg$-admissible strategies is defined by 
\begin{equation}
\label{admissible}
\Ac(\gg):=\bigcup_{M\in \Mo(\G_0)}\Ac^M(\gg).
\end{equation}
\edefi

Beside dynamic trading we allow for static trading in options $\Xc$, which results in semi--static trading as defined below.

\bdefi
\label{def:semitrading}
(i) Let $M\in\Mo(\G_0)$. A $(\gg,M)$-admissible semi--static strategy is a pair $(\mo, \gamma)$ where $\gamma\in \Ac^M(\gg)$ and $\mo=(\mo^\no)_{\no\in\No}$ such that for each $\no\in\No$, $\mo^\no\in\Mo(\G_{-1})$ and for each atom of $\G_{-1}$, $A^\omega$, there exists a finite subset $\bet^\omega\subset \No$ such that $\mo^\no(\omega)=0$  for each $\no\notin\bet^\omega$. \\
(ii) The set of all $(\gg,M)$-admissible semi--static strategies is denoted by $\Ac^M_{\Xc}(\gg)$ and all $\gg$-admissible semi--static strategies are given by 
\begin{equation}
\Ac_\Xc(\gg):=\bigcup_{M\in \Mo(\G_0)}\Ac^M_\Xc(\gg).
\end{equation}
\edefi

Let $A^\omega$ denote an atom of $\G_{-1}$ and $(\mo, \gamma)$ be a $\gg$-admissible trading strategy. 
Note that, for each $\wt\omega\in A^\omega$, the initial cost of $(\mo,\gamma)$ equals 
$$\mo(\omega)\mathcal P(\Xc)=\mo(\wt \omega)\mathcal P(\Xc)=\mo^0(\wt \omega)+\sum_{\no\in\No}\mo^\no(\wt \omega)\Pc(\Xc^\no)=\mo^0(\wt \omega)+\sum_{\no\in\No_0^\omega}\mo^\no(\wt \omega)\Pc(\Xc^\no).$$ 
For each $\omega=(\omega^{(1)},...,\omega^{(d)},\omega^{(d+1)},...,\omega^{(d+K)})\in\I$ the final payoff of $(\mo,\gamma)$ is given by
\be
\label{self}
(\gamma \is S)_T(\omega) + (\mo\Xc)(\omega)=\int_0^T \gamma(\omega)_udS_u(\omega) + \sum_{\no\in\No} \mo^{\no}(\omega) \Xc^{\no}(\omega^{(1)}, ..., \omega^{(d)}).
\ee

\section{Pathspace approach to information quantification}
\label{s:ph}

We are now in a position to define the main quantities of interest: the robust pricing and hedging prices of an option. 
We work with a general filtration $\gg$ and this induces further difficulties, as compared with the case of the natural filtration $\ff$. 

We start with superhedging problem.
\bdefi\label{def:superhedge}
Let $A\subset \I$. The $\gg$-superhedging cost of $\xi$ on $A$ is given by
\begin{align*}
\vggg[A]{\xi}(\omega):=\inf\{&\mo(\omega)\Pc(\Xc):\,\exists \,(\mo, \gamma) \in\mathcal A_\mathcal X(\gg)\,\, \textrm{such that}\,\, \\
&\quad (\gamma \is S)_T(\wt\omega) + (\mo\Xc)(\wt \omega)\geq \xi(\wt\omega) \textrm{  for each  } \wt\omega\in A\},\quad \omega\in A,
\end{align*}
where $(\gamma \is S)_T + (\mo\Xc)$ is defined in \eqref{self}.
\edefi
Thus the $\gg$-superhedging cost of $\xi$ on $A$ is the pathwise infimum over all initial costs $\mo\Pc(\Xc)$ of $\gg$-admissible semi--static strategies $(\mo, \gamma) \in\mathcal A_\mathcal X(\gg)$ which super--replicate $\xi$ on $A$ over $[0,T]$ i.e., $\mo\Xc+(\gamma\is S)_T\geq \xi$ on $A$. 
We have the following obvious inequalities:
$$\vgg[A]{\xi}\leq \vggg[A]{\xi}\leq \vg[A]{\xi}\leq \vf[A]{\xi}.$$
As soon as $\G_{-1}$ is not trivial, $\mo\Pc(\Xc)$ is random and hence $\vggg[A]{\xi}$ is random as well and its measurability is not clear \emph{a priori}. However, the following result shows that the superhedging cost is constant on the atoms of $\G_{-1}$. In particular, when  $\G_{-1}$ has at most countably many atoms, e.g.\ is generated by a discrete random variable, then it follows that the superhedging cost is $\G_{-1}$-measurable.

\begin{Proposition}
\label{hedge}
The $\gg$-superhedging cost of $\xi$ on $\I$, defined in Definition \ref{def:superhedge}, is constant on atoms of $\G_{-1}$. Specifically, for any $\omega\in \I$ we have
$$
\vggg{\xi}(\omega) =\vggg{\xi}(\omega') =\vggg[A^{\omega}]{\xi}(\omega')\quad \forall \omega'\in A^\omega,
$$
where $A^{\omega}$ denotes the $\G_{-1}$-atom containing $\omega$. 
\end{Proposition}
\proof
Fix $\omega\in \I$ and $\omega'\in A^\omega$. 
It follows from $\mo^\no\in\Mo(\G_{-1})$ for each $\no\in\No$ that $\mo(\omega)\Pc(\Xc)=\mo(\omega')\Pc(\Xc)$ for any $\gg$-admissible strategy $(\mo, \gamma) \in\mathcal A_\mathcal X(\gg)$ which super--replicates $\xi$ on $A^\omega$. This in turn implies that
$$ \vggg[A^{\omega}]{\xi}(\omega)=\vggg[A^{\omega}]{\xi}(\omega').$$
It remains to argue that these are also equal to $\vggg{\xi}(\omega')$. Clearly the latter can only be larger as the super--hedging is required on a larger set. 
As for the reverse inequality, note that any $\gg$-admissible semi--static strategy $(\mo, \gamma)$ which super--replicates $\xi$ on $A^{\omega}$ can be extended to a strategy $(\bar \mo, \bar \gamma)$ super--replicating on $\I$ by taking $\bar \mo=\mo$ on ${A^{\omega}}$ and $\bar \mo^0=\norm{\xi}$ otherwise, $\bar \gamma=\gamma\id_{A^{\omega}}$.
\finproof
With a slight abuse of notation, we write $\vggg[A^{\omega}]{\xi}$ both for the constant function, extended to $\Omega$, as well as its value equal to $\vggg{\xi}_{|A^\omega}$.

We now turn to the pricing problem. In the classical approach markets with no-arbitrage are modelled using martingale measures. We denote by $\Mcal^\gg$ probability measures on $(\I,\F_T)$ such that $S$ is a $\gg$--martingale. For a given $A\in \F_T$, we look at possible classical market models which calibrate to market prices of options and are supported on $A$:
$$\Mcal^\gg_{\mathcal X, \mathcal P, A}:=\left\{\P\in \Mcal^\gg: \P(A)=1 \textrm{ and }\E_\P[\Xc^\no|\G_{-1}]=\mathcal P(\Xc^\no) \textrm{ for all }\no\in \No,\ \P\textrm{-a.s.}\right\}.$$
We emphasize that any measure $\P\in \Mcal^\gg_{\mathcal X, \mathcal P, A}$ is calibrated to the initial prices of options in $\Xc$. In particular, for any atom $A^\omega$ of $\G_{-1}$ one has $\E_\P[\Xc^\no\id_{A^\omega}]=\E_\P[\mathcal P(\Xc^\no)\id_{A^\omega}]$ for each $\no\in \No$. 

To consider the pricing problem we need to look at 
"$\sup_{\P\in\mathcal M^\gg_{\mathcal X, \mathcal P, \I}}\E_\P[\xi|\G_{-1}]$". However, unless $\G_{-1}$ is a trivial $\sigma$-field, the conditional expectation $\E_\P[\xi|\G_{-1}]$ is a random variable which is determined only $\P$-a.s. 
As measures in the set $\mathcal M^\gg_{\mathcal X, \mathcal P, \I}$ may be mutually singular we have to be careful about choosing a good version of these conditional expectations.

\begin{Lemma}
\label{rcp}
Let $\P\in \mathcal M^\gg_{\mathcal X, \mathcal P, \I}$.
Then, there exists a set $\Omega^\P\in \G_{-1}$ with $\P(\Omega^\P)=1$ and a version $\{\P_{\omega}\}$ of the regular conditional probabilities of $\P$ with respect to $\G_{-1}$ such that for each $\omega \in \Omega^\P$, $\P_{\omega}\in \mathcal M^\gg_{\mathcal X, \mathcal P, A^{\omega}}$ where $A^\omega$ is the $\G_{-1}$-atom containing $\omega$.
\end{Lemma}
The proof, based on standard existence results for regular conditional probabilities, is reported in the Appendix. Suppose now that we fix a representation $\Omega^\P$ in Lemma \ref{rcp} for each $\P\in M^\gg_{\mathcal X, \mathcal P, \I}$ and use these to define the market model price as follows:
\bdefi\label{def:marketprice}
Let $A\in \F_T$. The $\gg$-market model price of $\xi$ on $A$ is defined by 
$$\pggg[A]{\xi}(\omega):=\sup_{\P\in \mathcal M^\gg_{\mathcal X, \mathcal P, A}}\bar\E_{\P_\omega}[\xi],\quad
\omega\in \I,$$
where $\bar\E_{\P_\omega}[\xi]=\E_{\P_\omega}[\xi]$ for $\omega\in \Omega^\P$ and $\bar\E_{\P_\omega}[\xi]=-\infty$ for $\omega \in \Omega\backslash \Omega^\P$, where $\Omega^\P$ is a set from Lemma \ref{rcp}.
\edefi
We note the following obvious inequalities:
$$\pgg[A]{\xi}\leq P^{\gg}_{\mathcal X, \mathcal P, A}(\xi)\leq\pg[A]{\xi}\leq \pf[A]{\xi}.$$
We now show that, as in the case of superhedging, the model price is constant on atoms of $\G_{-1}$ and is uniquely defined, irrespective of the choice of $\Omega^\P$ above. It follows that, while in general the measurability of $\pggg{\xi}$  is not clear, when $\G_{-1}$ has at most countably many atoms then $\pggg{\xi}$ is $\G_{-1}$-measurable.
\begin{Proposition}
\label{price}
The $\gg$-market model price of $\xi$ on $\I$ is uniquely defined on $\I$ and is constant on atoms of $\G_{-1}$. Specifically, for $\omega\in \I$ we have
$$\pggg{\xi}(\omega)=\pggg{\xi}(\omega')=\pggg[A^\omega]{\xi}(\omega')=\sup_{\P\in \mathcal M^\gg_{\mathcal X, \mathcal P, A^\omega}}\E_{\P}[\xi],\quad \forall \omega'\in A^\omega,
$$ 
where $A^\omega$ is the $\G_{-1}$-atom containing $\omega$.
\end{Proposition}
\proof
Note that if $\P\in \mathcal M^\gg_{\mathcal X, \mathcal P, A^\omega}$, then $A^\omega\subset \Omega^\P$ and $\P_{\omega'}=\P$ for all $\omega'\in A^\omega$. In particular, $\pggg[A^\omega]{\xi}(\omega')=\sup_{\P\in \mathcal M^\gg_{\mathcal X, \mathcal P, A^\omega}}\E_{\P}[\xi]$ for all $\omega'\in A^\omega$, i.e.\ $\pggg[A^\omega]{\xi}$ is well defined and constant on $A^\omega$. Further, the obvious inclusion $\mathcal M^\gg_{\mathcal X, \mathcal P, A^\omega}\subset \mathcal M^\gg_{\mathcal X, \mathcal P, \I}$ combined with Lemma \ref{rcp}, show that the set $\mathcal M^\gg_{\mathcal X, \mathcal P, A^\omega}$ is composed of conditional probabilities of measures in $\mathcal M^\gg_{\mathcal X, \mathcal P, \I}$. In consequence,  
$$\pggg{\xi}(\omega')=\pggg[A^\omega]{\xi}(\omega'), \quad \omega'\in A^\omega$$ as required.
In particular, $\pggg{\xi}$ is well defined and constant on atoms and does not depend on the choice of $\Omega^\P$ for $\P \in \mathcal M^\gg_{\mathcal X, \mathcal P, \Omega}$. Finally, if for some $\omega\in \I$, $\mathcal M^\gg_{\mathcal X, \mathcal P, A^\omega}=\emptyset$ then, for any $\P\in \mathcal M^\gg_{\mathcal X, \mathcal P, \I}$, $\Omega^\P\cap A^\omega=\emptyset$ and the definitions agree giving 
$$\mathcal M^\gg_{\mathcal X, \mathcal P, \Omega}(\omega)=-\infty.$$
\finproof
With a slight abuse of notation, we write $\pggg[A^\omega]{\xi}$ both for the constant function, extended to $\Omega$, as well as its value equal to $\sup_{\P\in \mathcal M^\gg_{\mathcal X, \mathcal P, A^\omega}}\E_{\P}[\xi]=\pggg{\xi}_{|A^\omega}$.
\brem
We note that Proposition \ref{price} implies that when computing the $\gg$-market model price of $\xi$ on $\I$, or on some set in $\G_{-1}$, it is sufficient to maximise over measures supported on a single atom of $\G_{-1}$. Such measures are not necessarily the extreme measures in $\mathcal M^\gg_{\mathcal X, \mathcal P, \Omega}$, which were the focus in \cite{acciaio2015semi}. 
\erem

\brem 
In this paper we are interested in information which may affect pricing and hedging. Other, ``irrelevant'', information can be modelled via enlargement of measurable space $(\I,\F_T)$ in the following way. Let $\wt \Omega=\Omega\times \Omega{'}$ where $(\Omega^{'}, \G)$ is another measurable space, and denote $\wt \omega:=(\omega, u)\in \Omega\times \Omega{'}$. Assume that additional information is provided by $Z:\wt \Omega\to\R$ satisfying $Z(\wt \omega)=Z(u)$.
As before we may considered $\gg=\ff\lor \sigma(Z)$, where filtration $\ff$ can be naturally interpreted on $(\wt\Omega, \F_T\otimes\G)$.
Then, any atom $A^{(\omega,u)}$ of $\G_0$ (resp. $\G_{-1}$) is equal to (resp. contained in) $\I\times \{v: Z(v)=Z(u)\}$. In particular, $Z$ does not affect the martingale condition, $\mathcal M(\gg)=\mathcal M(\ff)$
and $\gamma^\ff=\gamma^\gg(c)$ for any $c\in Z(\Omega)$ etc., and the pricing problem is unchanged by the addition of the information carried by $Z$. Such situation was considered in the $\mathcal Q$-quasi-sure reduced form modelling in Biagini \cite{biagini2017}, where $(\Omega^{'},\G)=([0,1],\mathcal B([0,1]))$ by imposing $\wt {\mathcal Q}=\{\wt \Q:=\Q\times \mathcal U([0,1]) : \Q\in\mathcal Q\} $.
\erem

\section{Pricing--hedging duality under initially enlarged filtration}
\label{s:duality} 

Our first contribution in this paper, developed in Propositions \ref{hedge} and \ref{price} above, was to describe the superhedging cost and the pricing problem for an agent with additional information. We turn now to our second main contribution: understanding when pricing--hedging duality for a regular agent carries over to the informed one? It is straightforward to see that an inequality holds in general:
\begin{Lemma}
\label{classical}
The $\gg$-superhedging cost $V^{\gg}_{\mathcal X, \mathcal P, \I}(\xi)$ and the $\gg$-market model price $P^{\gg}_{\mathcal X, \mathcal P, \I}(\xi)$ of $\xi$ on $\Omega$ satisfy
$$V^{\gg}_{\mathcal X, \mathcal P, \I}(\xi)(\omega)\geq P^{\gg}_{\mathcal X, \mathcal P, \I}(\xi)(\omega)\quad \forall \omega\in \Omega \;.$$
\end{Lemma}
Thanks to the results of Section \ref{s:ph}, the proof is carried out on atoms of $\G_{-1}$ and is reported in Appendix. Our goal in this section is to provide sufficient conditions for equality in the above inequality. We start with the case when $\gg$ is an initial enlargement of $\ff$. Then, $\gamma$ is $\gg$-progressively measurable if and only if
$\gamma(\omega)_t=\gamma(\wt \omega)_t$ whenever $\omega\vert_{[0,t]}=\wt \omega\vert_{[0,t]}$ and $\omega \sim_{\G_0} \wt \omega$.

\subsection{Preliminaries: pricing--hedging duality with beliefs}
We start by recalling notions and results from Hou \& Ob\l\'{o}j \cite{ho_beliefs}. As explained before, we use their setup with beliefs to \emph{zoom in} on the part of the pathspace considered by an informed agent. 
For $\mathfrak{P}\in \F_T$ let $\tvf[\mathfrak P]{\xi}$ be the approximate $\ff$-superhedging cost of $\xi$ on $\mathfrak P$, i.e., 
\begin{align*}
\tvf[\mathfrak P]{\xi}:=\inf\{&\mo\mathcal P(\Xc):\quad \exists (\mo, \gamma) \in\mathcal A_\mathcal X(\ff)\,\, \textrm{such that}\,\,\\
&  (\gamma\circ S)_T(\omega)+(\mo\Xc)(\omega)\geq \xi(\omega) \;\; \forall\; \omega\in \mathfrak P(\varepsilon) \quad \textrm{for some}\quad \varepsilon >0\},
\end{align*}
where 
\begin{equation}
\label{ball}
\mathfrak P(\varepsilon)=\{\omega \in \I: \inf_{\wt \omega \in \mathfrak P}\norm{\omega-\wt \omega}\leq \varepsilon\}.
\end{equation}

Similarly, let $\tpf[\mathfrak P]{\xi}$ be the approximate $\ff$-market model price of $\xi$, i.e., 
$$\tpf[\mathfrak P]{\xi}:=\lim_{\eta \searrow 0 } \sup_{\P\in \mathcal M^{\ff, \eta}_{\mathcal X, \mathcal P, \mathfrak P}}\E_\P[\xi]\quad
$$
with $\mathcal M^{\ff, \eta}_{\mathcal X, \mathcal P, \mathfrak P}:=\{\P\in \mathcal M^\ff_\Omega$ : $\P(\mathfrak P(\eta))>1-\eta$ and $|\E_\P[\Xc^\no]-\mathcal P(\Xc^\no)|<\eta$ for all $\no\in \No$ $\}$. \\
The following assumption says that the vector $\Xc$ is not too large, and initial prices of dynamically traded options are not "on the boundary of no-arbitrage region".
\begin{ass}
\label{chi}
(i) $Lin_1(\mathcal X)$ is a compact subset of $C(\Omega, \mathbb R)$, where $Lin_N(\mathcal X)$ is given by
\begin{equation}
\label{linN}
\left\{\sum_{\no\in\No}\mo^\no\Xc^\no: \mo=(\mo^\no)_{\no\in\No}\in\R^\No \,\,\textrm{with finitely many $\mo^\no\neq 0$ and } \,\sum_{\no\in\No}|\mo^\no|\leq N\right\}.\end{equation}
(ii) Either $K=0$ or there exists an $\varepsilon >0$ 
such that for any $(p_k)_{k\leq K}$ with $|\mathcal P(X^{(c)}_k)-p_k|\leq \varepsilon$ for all $k\leq K$, $\mathcal M^{\ff}_{\wt{ \I}}\neq \emptyset$ where 
$$
\wt {\I}=\{\omega\in \I : S^{(d+i)}_T(\omega)=X^{(c)}_i(\omega)/p_i\quad \forall\; i\leq K\}.
$$
\end{ass}

\bethe[Hou \& Ob\l\'oj \cite{ho_beliefs}]
\label{ho1} Suppose that Assumption \ref{chi} holds.
Let $\frak P\in \F_T$ be such that $\mathcal M^{\ff, \eta}_{\mathcal X, \mathcal P, \mathfrak P}\neq \emptyset$ for any $\eta>0$.  
Then, for any uniformly continuous and bounded $\xi$, the approximate pricing--hedging duality holds:
$$\tvf[\mathfrak P]{\xi}= \tpf[\mathfrak P]{\xi}.$$
\eethe

Let us consider a general filtration $\gg$ and assume that $\mathcal M^\gg_{\mathcal X, \mathcal P, \Omega}\neq \emptyset$. The above results on $\I$ without options (i.e., $P^\gg_\I(\xi)=V^\gg_\I(\xi)$) extend to the case with static hedging in options by applying min--max type of argument exactly in the same spirit as in Sections 4.1 and 4.2 of \cite{ho_beliefs}. Hence, as a corollary of the proof of Theorem 4.3 in \cite{ho_beliefs} we obtain the following result.

\bcor
\label{addoptions}
Let $\gg$ be an \emph{arbitrary} filtration such that $\G_{-1}$ is a trivial $\sigma$-field. 
Suppose that Assumption \ref{chi} holds and that $\mathcal M^\gg_{\mathcal X, \mathcal P, \Omega}\neq \emptyset$.
Moreover assume that there is pricing--hedging duality for the model without options, i.e., 
$P^\gg_\Omega(\xi)=V^\gg_\Omega(\xi)$ for all uniformly continuous and bounded $\xi$.
Then 
$$
\wt P^\gg_{\mathcal X, \mathcal P, \Omega}(\xi)=\wt V^\gg_{\mathcal X, \mathcal P, \Omega}(\xi)=V^\gg_{\mathcal X, \mathcal P, \Omega}(\xi) \quad \textrm{ for all uniformly continuous and bounded $\xi$.}
$$
\ecor

\subsection{Duality in an enlarged filtration: case of $\gg^+$} 
From Theorem \ref{ho1}, taking $\frak P$ to be atoms in $\G^+_{-1}=\G_0$, we can deduce a pricing--hedging duality in the enlarged filtration. We isolate the following assumption which is often invoked. 
\begin{ass}
\label{fduality}
The set $\frak P \in \F_T$ is such that $\mathcal M^\ff_{\mathcal X, \mathcal P, \frak P}\neq \emptyset$, and 
\begin{equation}
\label{as}
\pf[\frak P]{\xi}=
\vf[\frak P]{\xi}
\quad \textrm{for any bounded uniformly continuous $\xi$}.
\end{equation}
\end{ass}

\bethe
\label{duality}
Suppose Assumption \ref{chi} holds and let $Z$ be a random variable and $\gg:=\ff\lor \sigma(Z)$. 
Assume that for each value $c\in Z(\Omega)$ either $\{Z=c\}$ satisfies Assumption \ref{fduality} or  $\mathcal M^\ff_{\mathcal X, \mathcal P, \{Z=c\}}= \emptyset$. 
Then, there is no duality gap in $\gg^+$ on the set $\{\omega: \mathcal M^\ff_{\mathcal X, \mathcal P, \{Z=Z(\omega)\}}\neq \emptyset\}$, i.e.,  
$\vgg{\xi}(\omega)=\pgg{\xi}(\omega)$ holds for any $\omega$ such that $\mathcal M^\ff_{\mathcal X, \mathcal P, \{Z=Z(\omega)\}}\neq \emptyset$ and any bounded uniformly continuous $\xi$. 
\eethe
\proof
In the first step we prove that for each value $c\in Z(\Omega)$ such that $\mathcal M^\ff_{\mathcal X, \mathcal P, \{Z=c\}}\neq \emptyset$ we have
\begin{equation}
\label{ee}
\pf[\{Z=c\}]{\xi}=\pgg[\{Z=c\}]{\xi}\leq \vgg[\{Z=c\}]{\xi}= \vf[\{Z=c\}]{\xi}.
\end{equation}
First let us prove the last equality. To this end note that any $\gg^+$-progressively measurable process $\gamma$ is of the form $\gamma(\omega)_t=\Gamma(Z(\omega), t, \omega)$ where for each $t\in[0,T]$ the mapping $\Gamma: \R\times [0,t]\times \Omega \to \R^{d+K}_+$ is $\mathcal B(\R)\otimes \mathcal B([0,t])\otimes \F_s $-measurable. 
So there exists $\ff$-progressively measurable process $\gamma^\ff$ given by $\gamma^\ff(t, \omega)=\Gamma(c, t, \omega)$ which is equal to $\gamma$ on $\{Z=c\}$ thus the last equality in \eqref{ee} follows.\\ 
To show the first equality, it suffices to show that $\mathcal M^{\gg^+}_{\mathcal X, \mathcal P, \{Z=c\}}=\mathcal M^\ff_{\mathcal X, \mathcal P, \{Z=c\}}$. For any $\P\in \mathcal M^\ff_{\mathcal X, \mathcal P, \{Z=c\}}$ and $0\leq s\leq t\leq T$ we have
$$
\E_\P[S_t|\G^+_s]=\E_\P[S_t|\F_s\lor\sigma(Z)]=\E_\P[\E_\P[S_t|\F^\P_s]|\F_s\lor\sigma(Z)]=\E_\P[S_s|\F_s\lor\sigma(Z)]=S_s,
$$
where $\F^\P_s$ is a $\P$-completion of $\F_s$, showing $\P\in \mathcal M^{\gg^+}_{\mathcal X, \mathcal P, \{Z=c\}}$. The reverse inclusion is clear. Finally,  the middle inequality in \eqref{ee} is implied by Lemma \ref{classical}. 

By representations given in Propositions \ref{hedge} and \ref{price}, the proof is completed since
 we get that $\vgg[\{Z=c\}]{\xi}=\pgg[\{Z=c\}]{\xi}$ for any value $c\in Z(\Omega)$ for which $\mathcal M^\ff_{\mathcal X, \mathcal P, \{Z=c\}}\neq \emptyset$ and any bounded uniformly continuous claim $\xi$.
\finproof

\begin{rem}
\label{ppt}
Under Assumption \ref{chi}, thanks to Theorem \ref{ho1}, we have
\begin{equation}
\label{ee2}
\tvf[\{Z=c\}]{\xi}=\tpf[\{Z=c\}]{\xi}.
\end{equation}
Combining equality \eqref{ee2} with the following general inequalities
$$
\pf[\{Z=c\}]{\xi}\leq \vf[\{Z=c\}]{\xi}\leq \tvf[\{Z=c\}]{\xi}
$$
we conclude that in Theorem \ref{duality}, instead of assuming that \eqref{as} holds, it is enough to assume that
$$
\pf[\{Z=c\}]{\xi}= \tpf[\{Z=c\}]{\xi}.
$$
\end{rem}

We present now two examples where the informed agent has additional knowledge about the price process at the terminal date. 
These corresponds to two very different situations: first when the informed agent gets rather detailed information and there is a continuum of atoms, second when the informed agent gets binary information (and $\G_{-1}$ only has two atoms). In both cases we only consider the situation with no static trading: $\No=\{0\}$ and such that pricing--hedging duality in $\gg^+$ holds.

\begin{exam}
\label{s1}
Consider a one-dimensional setting with no statically or dynamically traded options: $d=1$, $K=0$ and $\No=\{0\}$. 
The informed agent acquires detailed knowledge of the stock price process: namely she knows the maximum deviation over time interval $[0,T]$ of the stock price from the initial price. 
Naturally, the agent does not know the sign of the deviation as this would give an instant arbitrage. 
This situation corresponds to taking  $Z=\sup_{t\in [0,T]}|S_t-1|$. 
\\
For $c> 1$, the market model price $P^{\gg^+}_{\{Z=c\}}(\xi)=-\infty$ as any martingale measure $\P\in \mathcal M_{\{Z=c\}}^{\gg^+}$ would satisfy 
$$1=\P\left (\sup_{t\in[0,T]}S_t\geq 1+c\right )\leq \frac{1}{1+c}<1,$$
thus the set of martingale measures $\mathcal M_{\{Z=c\}}^{\gg^+}$ must be empty. 
Likewise the super--hedging cost $V^{\gg^+}_{\{Z=c\}}(\xi)=-\infty$ as a long position in the stock generates arbitrage.\\
Fix $c\leq 1$. Our proof is similar to Example 3.12 in \cite{ho_beliefs}. 
Firstly observe that under any $\P\in\mathcal M^{\gg^+}_{ \{Z=c\}}$, $S$ is a uniformly integrable martingale with $S=S^{\tau_c}$ $\P$-a.s. where $\tau_{c}:=\inf\{t: S_t\in \{1+c, 1-c\}\}$.

For each $N$ there exists a measure $\P^{N}\in {\mathcal M}_{\{Z=c\}}^{\ff, 1/N}$ such that
$$\E_{\P^{N}}[\xi]\geq \sup_{\P\in {\mathcal M}_{\{Z=c\}}^{\ff,1/N}}\E_\P[\xi]-\frac{1}{N}.$$
Since  $\P^N$ is a martingale measure for $S$, Doob's martingale inequality implies:
$$\P^{N}(\norm{S}>M)\leq \frac{1}{M},$$
and then, defining $\tau_M:=\inf\{t: S_t=M\}$ for an arbitrary large $M$, leads to:
\begin{align*}
|\E_{\P^{N}}[\xi(S)-\xi(S^{\tau_M})]|
&\leq\frac{2\norm{\xi}}{M}.
\end{align*}
Let $\pi^{N}$ be the distribution of $S^{\tau_M}_T$ under $\P^N$, for $N\in \mathbb N$.  
Since $\pi^N([0,M])=1$ for each $N$, this is a tight family of probability measures and therefore a converging subsequence $(\pi^{N_k})_k$ exists. 
Denote the limit of $(\pi^{N_k})_k$ by $\pi$ and note that, since each measure $\pi^N$ has mean equal to one so does $\pi$. \\
For each $\varepsilon>0$, weak convergence of measures and Portemanteau Theorem imply that
\begin{align}
\label{porte}
\pi([1-c-\varepsilon, 1+c+\varepsilon])
\geq \limsup_{k\to \infty}\pi^{N_k}([1-c-\varepsilon, 1+c+\varepsilon])
=1
\end{align}
since, letting $U_N:=[1-c-\frac{1}{N}, 1+c+\frac{1}{N}],$ one has
\begin{align*}
\P^N(S^{\tau_M}_T\in U_N )
=\P^N(\norm{S}\leq M, \,\,S_T\in U_N)
\geq \P^N(\norm{S}\leq c+1, \,\,S_T\in U_N)
\geq 1-\frac{1}{N}.
\end{align*}
It holds for each $\varepsilon>0$ hence $\pi([1-c, 1+c])=1$. 
Finally,  
\begin{align*}
\wt P^\ff_{\{Z=c\}}
&=\lim_{N\to\infty}\sup_{\P\in \underline{\mathcal M}_{\{Z=c\}}^{\ff,1/N}}\E_\P[\xi]\leq \lim_{N\to\infty} \left( \E_{\P^{N}}[\xi]+\frac{1}{N}\right)\\
&\leq \limsup_{k\to \infty} \sup_{\P\in \underline{\mathcal M}^\ff \; \textrm{s.t.}\; \mathcal L(S_T)=\pi^{N_k}}\E_\P[\xi]+\frac{2|\!|\xi|\!|}{M}\\
&\leq \sup_{\P\in \underline{\mathcal M}^\ff \; \textrm{s.t.}\; \mathcal L(S_T)=\pi}\E_\P[\xi]+\frac{2 |\!|\xi|\!|}{M}
\leq \sup_{\P\in \underline{\mathcal M}^\ff_{\{Z=c\}}}\E_\P[\xi]+\frac{2d |\!|\xi|\!|}{M}
\end{align*}
where the fourth inequality holds by Lemma 4.4 in \cite{ho_beliefs} and the fifth {since under any $\P\in \mathcal{M}^\ff_{\{Z=c\}}$ the distribution of $S_T$ equals $\pi$}.
Since $M$ was arbitrary we obtain that $\wt P^\ff_{\{Z=c\}}\leq P^\ff_{\{Z=c\}}$ and by Theorem \ref{duality} and Remark \ref{ppt} we conclude that there is no duality gap in $\gg^+$.
\end{exam}

\begin{exam}
\label{s2}
Assume $\mathcal X=\emptyset$, $d=1$ and $K=0$. Take $Z=\id_{\{S_t\in(a,b) \; \forall t\in [0,T]\}}$ where $a<1<b$. 
We use Theorem \ref{duality} to show that the pricing--hedging duality holds in $\gg^+$. \\
Equality $P^\ff_{\{Z=0\}}(\xi)=\wt P^\ff_{\{Z=0\}}(\xi)$ follows by the same arguments used in the previous example and the proof of Example 3.12 in \cite{ho_beliefs} since the set $[0,a]\cup[b,M]$ is compact for any $M$. 
The only difference lies in arguing an analogous inequality to \eqref{porte}. Here we have
\begin{align*}
\pi([0, a+\varepsilon]\cup[b-\varepsilon, \infty))
\geq \limsup_{k\to \infty}\pi^{N_k}([0, a+\varepsilon]\cup[b-\varepsilon, \infty))
=1
\end{align*}
since
\begin{align*}
\P^N(S^{\tau_M}_T\in [0, a+1/N]\cup[b-1/N, \infty) )
\geq \P^N(S_T\in [0, a+1/N]\cup[b-1/N, \infty))
\geq 1-\frac{1}{N}.
\end{align*}
It holds for each $\varepsilon>0$ and hence $\pi([0,a]\cup[b,\infty))=1$. 

To prove that $P^\ff_{\{Z=1\}}(\xi)=\wt P^\ff_{\{Z=1\}}(\xi)$ we first note that, again by the same type of arguments as in the proof of Example 3.12, for the interval $[a,b]$ we have 
$$P^\ff_{\{S_t\in [a,b]\; \forall t\in [0,T]\}}(\xi)=\wt P^\ff_{\{S_t\in [a,b]\; \forall t\in [0,T]\}}(\xi)=\wt P^\ff_{\{Z=1\}}.$$
 Thus, in order to show that $P^\ff_{\{Z=1\}}(\xi)=\wt P^\ff_{\{Z=1\}}(\xi)$ it is enough to prove that 
 $$P^\ff_{\{S_t\in [a,b]\;\forall t\in [0,T]\}}(\xi)=P^\ff_{\{Z=1\}}(\xi).$$ 
Take $\P\in \mathcal M^{\ff}_{\{S_t\in[a,b]\; \forall t\in [0,T]\}}$.
For each $k\in (0,1)$ define $\wt S^k$ as 
$$\wt S^k_t(\omega):=k\omega_t+(1-k)\quad \textrm{ for}\quad \omega\in \{S_t\in[a,b] \;\;\forall t\in[0,T]\}.$$
Then $$\P\circ (\wt S^k)^{-1}\in \mathcal M^{\ff}_{\{S_T\in(a,b)\}}\quad\textrm{and}\quad\norm {\wt S^k(\omega)-\omega}\leq (1-k)[(b-1)\lor(1-a)].$$
Thus $$|\E_\P[\xi]-\E_\P[\xi\circ \wt S^k ]|\leq e_\xi((1-k)[(b-1)\lor(1-a)])$$ and, since $k$ is arbitrary close to 1, there exists sequence of measures $\P^N\in \mathcal M^{\ff}_{\{S_T\in(a,b)\}}$ such that $\E_\P[\xi]=\lim_{N\to \infty}\E_{\P^N}[\xi]$.
\end{exam}

\subsection{Duality in an enlarged filtration: case of $\gg^-$}
We turn now to the pricing--hedging duality in $\gg^-$ in the case $\No=\{0\}$. We recall that $\G^-_{-1}=\{\emptyset, \I\}$ which models the situation when the static position in $\Xc^0$ has to be determined before acquiring any additional information. 

\begin{Theorem}
\label{gminus}
Assume that $\No=\{0\}$. 
Let $Z$ be a random variable such that for each $c\in Z(\Omega)$ the set $\{Z=c\}$ satisfies Assumption \ref{fduality}.
Define $\gg=\ff\lor \sigma(Z)$. 
Assume moreover that $\mathcal M^{\gg^-}_{\mathcal X, \mathcal P, \Omega}\neq \emptyset$. 
Then pricing--hedging duality holds in $\gg^-$: 
$$V^{\gg-}_\Omega(\xi)=P^{\gg-}_\Omega(\xi) \quad \forall \textrm{ bounded uniformly continuous }\xi.$$
\end{Theorem}
\proof
We will prove the following sequence of equalities
\begin{equation}
\begin{split}
V^{\gg^-}_\I(\xi)&=\sup_{c\in Z(\Omega)} V^{\gg^+}_{\{Z=c\}}(\xi)=\sup_{c\in Z(\Omega)} P^{\gg^+}_{\{Z=c\}}(\xi)\\
&=\sup_{\P\in\bigcup_{c\in Z(\Omega)} \mathcal M^{\gg^+}_{\{Z=c\}}}\E_\P[\xi]=\sup_{\P\in \mathcal M^{\gg^-}_\I}\E_\P[\xi] =P^{\gg^-}_\I.
\label{eq:chain_eq_2}
\end{split}
\end{equation}
Let us start with first equality. Since for each $c\in Z(\Omega)$, $V^{\gg^+}_{\{Z=c\}}(\xi)\leq  V^{\gg^-}_{\I}(\xi)$, we get that $\sup_{c\in Z(\Omega)} V^{\gg^+}_{\{Z=c\}}(\xi)\leq V^{\gg^-}_{\I}(\xi)$. 
To show the reverse inequality, fix $\varepsilon>0$. Then, for each $c\in Z(\I)$, there exists $\gamma^c\in\Ac(\gg)$
such that 
$$V^{\gg^+}_{\{Z=c\}}(\xi)+\varepsilon + (\gamma^c\is S)_T\geq \xi \quad \textrm{on $\{Z=c\}$}.$$
Define the strategy $\gamma$ as $\gamma(\omega)=\sum_{c\in Z(\Omega)}\gamma^c(\omega)\id_{\{Z(\omega)=c\}}$, which belongs to $\Ac(\gg)$ and satisfies
$$\sup_{c\in Z(\Omega)}V^{\gg^+}_{\{Z=c\}}(\xi)+\varepsilon + (\gamma\is S)_T\geq \xi \quad \textrm{on $\I$}.$$
Then, from the definition of super--hedging cost, we conclude that $$\sup_{c\in Z(\Omega)}V^{\gg^+}_{\{Z=c\}}(\xi)+\varepsilon\geq V^{\gg^-}_{\I}(\xi).$$
As $\varepsilon>0$ was arbitrary one has $\sup_{c\in Z(\Omega)} V^{\gg^+}_{\{Z=c\}}(\xi)\geq  V^{\gg^-}_{\I}(\xi)$. 

By our assumption $\{\omega: \mathcal M^\ff_{\{Z=Z(\omega)\}}\neq \emptyset\}=\I$ and pricing--hedging duality holds on $\{Z=c\}$ for each $c\in Z(\Omega)$. Thus the second equality in \eqref{eq:chain_eq_2} follows. 
Whereas the third and fifth equalities in \eqref{eq:chain_eq_2} hold by definition. 
To show the fourth one, note that, one inequality is immediate since for any $c$, $\Mcal^{\gg^+}_{\{Z=c\}}\subset \Mcal^{\gg^-}_{\I}$ and the other inequality then follows by Lemma \ref{rcp}.  
\finproof

\brem
We would like to emphasize that we do not have supremum representation as in \eqref{eq:chain_eq_2} neither for $V^{\gg-}_\Omega$ nor for $P^{\gg-}_\Omega$ when we consider the case of $\No\neq\{0\}$. The static hedging position may have opposite direction on different atoms and therefore simple aggregation is not possible. Similarly, calibration on each atom separately is much more restrictive condition than unconditional calibration.
\erem

\section{Timing of information arrival, dynamic programming principle and pricing--hedging duality}
\label{s:dpp}

To extend the initial enlargement perspective we study now the case where the additional information is disclosed at time $T_1\in (0,T)$, i.e.\ the filtration $\gg$ is of the form: $\G_t=\F_t$ for $t\in[0,T_1)$ and $\G_t=\F_t\lor \sigma(Z)$ for $t\in [T_1,T]$. 
We divide our problem into two time intervals using the results from the previous sections. 
First we look at the pricing and hedging problems on $[T_1, T]$ and then on $[0, T_1]$. 
Along this section we assume that there are no dynamically traded options, i.e., $K=0$.
Moreover we assume that $Z$ is of a special form, namely, $Z$ satisfies
\begin{align}
\label{similarZ}
Z(\omega)=
\begin{cases}
\wt Z\left (\frac{\omega\vert_{[T_1, T]}}{\omega_{T_1}}\right) & \textrm{if  } \;\; \omega_{T_1}>0\\
1& \textrm{if  }\;\; \omega_{T_1}=0
\end{cases}
\quad \textrm{for a r.v. $\wt Z$ on $\Omega\vert_{[T_1, T]}$.}
\end{align}
This condition encodes the idea that the additional information only pertains to the evolution of prices after time $T_1$ irrespectively of the prices on, or before, time $T_1$.

\subsection{Dynamic programming principle for $V^\gg$ and $P^\gg$}
We begin with two propositions where we develop the dynamic programming principle for the superhedging cost and the market model price. 
Note that the case $Z\equiv const$ and $\ff=\gg$ is also of interest, as it gives the dynamic programming principle under $\ff$.

\begin{Lemma}
\label{alfa_m}
Let $\Omega_{+}=\{\omega\in \Omega: \omega_{T_1}>0\}$. For each $v\in\Omega_{+}$ define the mapping $\ale^v$ on $\Omega_{+}\times\Omega_{+}$ with values in $\Omega_{+}$ by 
\begin{align}
\label{alfa}
\ale^{v}(\wt v, \omega):=
\begin{cases}
v\vert_{[0, T_1]}\otimes \frac{v_{T_1}}{\wt v_{T_1}}  \omega\vert_{[T_1, T]} & \omega \in B^{\wt v}\\
\wt v\vert_{[0, T_1]}\otimes \frac{\wt v_{T_1}}{v_{T_1}}   \omega\vert_{[T_1, T]} & \omega \in B^{v}\\
\omega & \omega \notin B^{v}\cup B^{\wt v}
\end{cases}
\end{align}
where $\lambda \omega$ is a multiplicative modification of $\omega$ by $\lambda$ in $\Omega$ and 
$v\vert_{[0, T_1]}\otimes \lambda \omega\vert_{[T_1, T]}$ means that the path is equal to $v$ on $[0,T_1]$ and to $\lambda \omega\vert$ on $[T_1, T]$.\\ 
Then, $\ale^v$ is $\F_T\otimes\F_T$-measurable.
\end{Lemma}
The proof simply exploits the defining properties of $\ale^v$ and is reported in Appendix. To formulate the dynamic programming principle for the superhedging cost we naturally extend the notions introduced in Definition \ref{def:admissibility} on the time interval $[0,T]$ to a subinterval $[T_1,T_2]\subset[0,T]$ and let $\Ac^M(\gg, [T_1,T_2])$ denote $(\gg,[T_1,T_2],M)$-admissible strategies  and $\Ac(\gg,[T_1,T_2])$ their union over $M\in \No(\G_0).$ 
Similarly, we Define the set of measures $\mathcal M^{\gg, [T_1,T]}_{A}$ concentrated on $A\in \F_T$ as follows:
$$\mathcal M^{\gg, [T_1, T]}_{A}:=\{\P: \textrm{$S$ is a $\gg$-martingale on $[T_1, T]$ and $\P(A)=1$}\}.$$
The following two results establish suitable regularity and dynamic programming principle for the superhedging cost and the pricing operator.
\begin{pro}
\label{dpV}
Let $B^\omega$ denote the $\F_{T_1}$-atom containing $\omega$. 
Then for a bounded uniformly continuous $\xi$ the following hold: \\
(i)  The mapping $V^{\gg, [T_1, T]}_\Omega(\xi): \Omega \to \mathbb R$ defined as 
$$V^{\gg, [T_1, T]}_\Omega(\xi)(\omega):=\inf\left\{x\in \R: \exists \,\,\gamma \in \mathcal A(\gg,[T_1, T])\;\; \textrm{such that} \; x+\int_{T_1}^T\gamma_t d S_t\geq \xi \; \textrm{on}\; B^\omega\right\}$$ 
is uniformly continuous and $\F_{T_1}$-measurable.\\
(ii) The dynamic programming principle holds in the form: 
$$V^{\gg, [0,T]}_\I(\xi)=V^{\ff, [0,T_1]}_\I\left(V^{\gg, [T_1, T]}_\Omega(\xi)\right).$$
\end{pro}
\begin{pro}
\label{dpP}
Let $B^\omega$ denote the $\F_{T_1}$-atom containing $\omega$ and assume that $\mathcal M^{\gg, [T_1, T]}_{B^\omega}\neq\emptyset$ for each $\omega$. 
Then for a bounded uniformly continuous $\xi$ the following hold. \\
(i) The mapping $P^{\gg, [T_1, T]}_\Omega(\xi)$ defined as 
$P^{\gg, [T_1, T]}_\Omega(\xi)(\omega):=\sup_{\P\in \mathcal M^{\gg, [T_1, T]}_{B^\omega}}\E_\P[\xi]$
is uniformly continuous and $\F_{T_1}$-measurable.\\
(ii) The dynamic programming principle holds in the form: 
$$P^{\gg, [0,T]}_\I(\xi)=P^{\ff, [0,T_1]}_\I(P^{\gg, [T_1, T]}_\Omega(\xi)).$$
\end{pro}

\begin{rem}
The dynamic programming principle for $P$ stated in Proposition \ref{dpP} (ii) is linked to conditional sublinear expectations studied in \cite[Theorem 2.3]{handel}. 
Since there is more structure in our set-up we prove it relying on uniform continuity of $\xi$ instead of a general analytic selection argument. 
\end{rem}

While both Propositions \ref{dpV} and \ref{dpP} seem natural their proofs are longer than one might expect and require certain technical details. We present them in Appendix. In particular, we note that Assumption \eqref{similarZ} is important in the proofs and one would not expect such results to hold for an arbitrary $Z$. Indeed, consider for exmaple $Z=|\ln S_T-\ln S_{T_1}|\id_{\{S_{T_1}=c\}\cap \{S_T>0\}}$, which violates \eqref{similarZ}. It is easy to see that in this case we can not guarantee the uniform continuity of $V^{\gg, [T_1, T]}_\Omega(\xi)$ or $P^{\gg, [T_1, T]}_\Omega(\xi)$.

\subsection{Pricing--hedging duality}
We show now that pricing-hedging duality holds for an agent with information flow $\gg$. This is done using analogous arguments to those in Sections 3 and 4 and treating independently each atom $B^\omega$ of $\F_{T_1}$. 
Firstly, as in Theorem \ref{duality} for the corresponding $\gg^+$ filtration, we look at the intersections with each level set of r.v. $Z$, 
namely at the sets $B^\omega\cap \{Z=c\}$ which form the atoms of $\G_{T_1}$. 
Secondly we aggregate over $Z$, as in Theorem \ref{gminus} for the corresponding $\gg^-$ filtration. 
The described operation reduces the problem to $[0,T_1]$ interval where $\gg$ coincides with $\ff$ where we conclude by dynamic programming principle.

\begin{Theorem}
\label{dp}
Assume that there are no options, i.e., $\No=\{0\}$. 
Let $Z$ be a random variable such that for each $c\in Z(\Omega)$ and each $\F_{T_1}$-atom $B^\omega$ the set $\{Z=c\}\cap B^\omega$ satisfies Assumption \ref{fduality} on $[T_1, T]$.
Assume moreover that $\mathcal M^{\gg}_{\Omega}\neq \emptyset$ and that Assumption \ref{chi} holds. 
Then for any bounded uniformly continuous $\xi$ 
$$V^{\gg, [T_1, T]}_\Omega(\xi)(\omega)=P^{\gg, [T_1, T]}_\Omega(\xi)(\omega) \quad \forall \omega \quad \textrm{and} \quad  V^{\gg}_{\I}(\xi)=P^{\gg}_{\I}(\xi).$$
\end{Theorem}
\proof 
We use the previous subsection to show the following equalities:
\begin{align*}
V^{\gg, [0,T]}_\I(\xi)&=V^{\ff, [0,T_1]}_\I(V^{\gg, [T_1, T]}_\Omega(\xi))=V^{\ff, [0,T_1]}_\I(P^{\gg, [T_1, T]}_\Omega(\xi))\\
&=P^{\ff, [0,T_1]}_\I(P^{\gg, [T_1, T]}_\Omega(\xi))=P^{\gg, [0,T]}_\I(\xi).
\end{align*}
After applying Propositions \ref{dpV} and \ref{dpP}, it remains to show the second and third equalities. 
To prove the second equality stating that $P^{\gg, [T_1, T]}_\Omega(\xi)=V^{\gg, [T_1, T]}_\Omega(\xi)$ firstly we remark some analogies with Section 3. Defining the $(\gg^+, [T_1, T])$-superhedging cost $V^{\gg^+,[T_1,T]}_{A}$ as
$$
V^{\gg^+,[T_1,T]}_{A}:=\inf\{x\in \G_{T_1}: \exists \;\gamma \in \mathcal A(\gg,[T_1, T]) \textrm{  such that  } x+\int_{T_1}^T\gamma_u dS_u\geq \xi \textrm{  on  } A\},
$$
analogously to Proposition \ref{hedge} we have that 

$$V^{\gg^+,[T_1,T]}_{B^{\omega}}(\omega')=V^{\gg^+,[T_1,T]}_{B^{ \omega}\cap \{Z=Z(\omega')\}}\quad \textrm{for each $\omega$ and each $\omega'\in B^\omega$.}$$ 
Analogously to Proposition \ref{price}, we deduce that 
$$P^{\gg^+,[T_1,T]}_{B^{\omega}}(\omega')=P^{\gg^+,[T_1,T]}_{B^{ \omega}\cap \{Z=Z(\omega')\}}\quad \textrm{for each $\omega$ and each $\omega'\in B^\omega$.}$$ 
Then, by mimicking the proof of Theorem \ref{gminus}, we pass from "$\gg^+$ to $\gg^-$", and derive that
$$V^{\gg, [T_1, T]}_\Omega(\xi)(\omega)=P^{\gg, [T_1, T]}_\Omega(\xi)(\omega) \quad \forall \omega .$$
The third equality follows by general duality result on $[0, T_1]$.\finproof

\begin{exam}
\label{d1}
Analogously to Example \ref{s1} let us look at the additional information in dynamic set-up which consists of a very detailed knowledge. 
We consider a one-dimensional setting with no statically or dynamically traded options: $d=1$, $K=0$ and $\No=\{0\}$. The informed agent acquires at time $T_1$ detailed knowledge of the stock prices process of the form  
$$Z=
\begin{cases}
\sup_{t\in [T_1, T]}|\ln S_t-\ln S_{T_1}|& S_t>0 \;\;\; \forall \; t\in[T_1,T]\\
 \,\, 1& \textrm{otherwise}
\end{cases}.$$
Note that for each $c\in Z(\Omega)$ and each $\F_{T_1}$-atom $B^\omega$, one has that
$$\mathcal M^{\gg, [T_1, T]}_{\{Z=c\}\cap B^\omega}=\mathcal M^{\ff, [T_1, T]}_{\{Z=c\}\cap B^\omega}\neq \emptyset$$
where the first equality holds since each $\{Z=c\}\cap B^\omega$ is a $\G_{T_1}$-atom. 
To show that 
$$ P^{\ff, [T_1, T]}_{\{Z=c\}\cap B^\omega}=\wt P^{\ff, [T_1, T]}_{\{Z=c\}\cap B^\omega}$$ 
we use the same arguments as in Example \ref{s1}.
Thus Assumption \ref{fduality} with no options ($\No=\{0\}$) on $[T_1, T]$ is satisfied for each each set $\{Z=c\}\cap B^\omega$.
Moreover note that, by concatenation of measure argument,
$
\mathcal M^{\gg}_{\Omega}\neq \emptyset
$
and, since $\No=\{0\}$, Assumption \ref{chi} is also satisfied.
We now apply Theorem \ref{dp} to show that there is no duality gap in $\gg$. 
\end{exam}

\begin{exam}
\label{d2}
Analogously to Example \ref{s2} we consider the additional information in dynamic set-up of binary type. 
The same as before we consider a one-dimensional setting with no statically or dynamically traded options: $d=1$, $K=0$ and $\No=\{0\}$. The informed agent acquires at time $T_1$ knowledge of  of the form 
$$Z=\id_{\left \{a<\frac{S_t}{S_{T_1}}<b \;\;\forall \;t\in[T_1,T]\right \}} \quad \textrm{ where } \quad a<1<b.$$
Along the lines of Example \ref{d1}, applying Theorem \ref{dp}, we deduce that there is no duality gap in $\gg$. 
\end{exam}

\subsection{The timing and value of information}
\label{sec:infovalue}

We explore now how an agent might value the additional information carried by $Z$ and if the timing of the arrival of such an information makes a difference. On one hand, in line with considering superhedging prices, our agent always considers the worst case to evaluate the added value of the information. On the other hand however, she is free to consider any payoff she wishes to best capture such added value. As highlighted above, our assumption \eqref{similarZ} encoded the idea that the additional information only pertains to the evolution of prices after time $T_1$ irrespectively of the prices on, or before, time $T_1$. As we show below, agent's flexibility to build custom payoff $\xi$ to best leverage the arriving information, means that she does not mind when the information arrives, as long as it arrives \emph{strictly} after time zero and before time $T$. If the information arrives at time $t=0$, the agent may not have time to enter into a suitable position to later exploit the arriving information and hence may associate lower value to such scenario.

To formalise the above discussion, let $Z: C([0,1])\to \mathbb R$ be a random variable. Then, for each $T_1\in[0,T)$, define a random variable 
$Z^{T_1}: C([0,T])\to \mathbb R$ by 
$$Z^{T_1}(\omega)=Z\left (\left(\frac{\omega_{t(T-T_1)+T_1}}{\omega_{T_1}} \right)_{t\in[0,1]}\right).$$
Such a $Z^{T_1}$ clearly satisfies condition \eqref{similarZ}. 
We denote by $\gg^{T_1}$ the corresponding enlargement of filtration, i.e., 
the filtration of the form $\G^{T_1}_t=\F_t$ for $t\in[0,T_1)$ and $\G^{T_1}_t=\F_t\lor \sigma(Z^{T_1})$ for $t\in [T_1,T]$. 

When hedging a payoff $\xi$ the agent considers the minimal advantage brought by the additional information:
\begin{equation}
\label{eq:v1}
v_{T_1}(Z;\xi):=\inf_{\omega\in\I}\left(V^{\ff}_\I(\xi)-V^{\gg^{T_1}}_\I(\xi)(\omega)\right)\qquad T_1\in[0,T),
\end{equation}
where the infimum over $\I$ only plays a role for $T_1=0$ since $V^{\gg^{T_1}}_\I(\xi)(\omega)=V^{\gg^{T_1}}_\I(\xi)$ is a constant for $T_1>0$.
To evaluate the robust advantage associated to the information $Z$ we need to normalise the payoffs. Let $\Phi$ denote the set of uniformly continuous functions $\xi:\I\to [0,1]$. Then the value of receiving information $Z$ at time $T_1$ for the superhedging problem is given by 
\begin{equation}
 \label{eq:value_info}
 v_{T_1}(Z):= \sup_{\xi\in \Phi} v_{T_1}(Z;\xi),\quad T_1\in [0,T).
\end{equation}

\begin{Theorem}
Let $Z: C([0,1])\to\R$ be given and assume that $\No=\{0\}$. Then:\\
(i) for any $T_1\in(0,T)$ and $\xi\in \Phi$ there exists $\wt \xi\in\Phi$ such that $v_{0}(Z;\xi)=v_{T_1}(Z;\wt \xi)$,\\
(ii) for any $T_1\in(0,T)$, $T^\prime_1\in(0,T)$ and $\xi\in\Phi$ there exists $\xi^\prime\in\Phi$ such that $v_{T_1}(Z;\xi)=v_{T^\prime_1}(Z;\xi^\prime)$.
\\
In consequence, 
$$v_0(Z)\leq v_{T_1}(Z)=v_{T_1^\prime}(Z),\quad 0<T_1<T_1^\prime<T.$$
\end{Theorem}
\proof
(i) For given $T_1\in(0,T)$ and $\xi$ let us define $\wt \xi$ by
$$\wt \xi(\omega):=\xi\left (\left(\frac{\omega_{t(T-T_1)/T+T_1}}{\omega_{T_1}}\right)_{t\in [0, T]}\right ),$$
where $\frac{0}{0}=1$. Then, by dynamic programming principle given in Proposition \ref{dpV}, 
$$V^\ff_\Omega(\wt \xi)=V^{\ff,[0,T_1]}_\Omega\left(V^{\ff,[T_1,T]}_\Omega(\wt \xi)\right)
=V^{\ff,[0,T_1]}_\Omega\left(V^{\ff}_\Omega(\xi)\right)=V^{\ff}_\Omega(\xi).$$
By analogous argument we also obtain that $V^{\gg^{T_1}}_\Omega(\wt \xi)=V^{\gg^{0-}}_\Omega(\xi).$ 
It remains to note that, by \eqref{eq:chain_eq_2}, $v^0(\xi)=V^\ff_\Omega(\xi)-V^{\gg^{0,-}}_\Omega(\xi)$.

(ii) For given $T_1\in(0,T)$, $T^\prime_1\in(0,T)$ and $\xi$ let us define $\xi^\prime\in \Phi$ by 
$\xi^\prime(\omega):=\xi(\omega_{\timech})$ where $\timech$ is the following time change
$$\timech(t):=t\frac{T_1}{T^\prime_1}\id_{\{t\in[0,T^\prime_1]\}}+\frac{T_1(T-t)+T(t-T^\prime_1)}{T-T^\prime_1}\id_{\{t\in(T^\prime_1, T]\}}.$$
Then, by Proposition \ref{dpV} and the form of additional information $Z$, it holds that $v^{T_1}(\xi)=v^{T^\prime_1}(\xi^\prime)$. 
\finproof

\appendix
\section{Proofs of Lemmas \ref{rcp}, \ref{classical} \& \ref{alfa_m}, and Propositions \ref{dpV} \& \ref{dpP} }
{\sc Proof of Lemma \ref{rcp}:}\\
Existence of $\P_\omega$ and its properties are all classical results, see Stroock \& Varadhan \cite[pp.~12--16]{strvar}.
Since, by Standing Assumption \ref{as:1}, $\G_{-1}=\sigma(B^{-1}_n,n\geq 1)$, there exists a set $\Omega_{-1}\in \G_{-1}$ such that  $\P(\Omega_{-1})=1$ and $\P_{\omega}(A^{\omega})=1$  for each $\omega\in \Omega_{-1}$.\\
Fix $t\geq s$ and $G\in \G_s$. Then, 
since $G_0:=\{\E_{\P_{\omega}}[(S_t-S_s)\id_G]>0\}\in \G_{-1}$, we get
\begin{align*}
0&=\E_\P[(S_t-S_s)\id_{G\cap G_0}]
=\E_\P[\E_\P[(S_t-S_s)\id_{G}|\G_{-1}]\id_{G_0}]
=\E_\P[\E_{\P_{\omega}}[(S_t-S_s)\id_{G}]\id_{G_0}],
\end{align*}
which implies that $\P$-a.s. $\E_{\P_{\omega}}[(S_t-S_s)\id_G]\leq 0$. In the same way we prove that $\P$-a.s. $\E_{\P_{ \omega}}[(S_t-S_s)\id_G]\geq 0$. 
So finally, $\P$-a.s. $\E_{\P_{\omega}}[(S_t-S_s)\id_G]= 0$ and therefore there exits a set $\Omega_{s,t,G}\in\G_{-1}$ such that $\P(\Omega_{s,t,G})=1$ and $\E_{\P_{\omega}}[(S_t-S_s)\id_G]= 0$ for each $\omega\in \Omega_{s,t,G}$.
To conclude that there exists a set $\Omega_m\in\G_{-1}$ such that $\P(\Omega_m)=1$ and $S$ is a $(\P_\omega,\gg)$-martingale for every $\omega\in \Omega_m$ we use continuity of paths of $S$ and Standing Assumption \ref{as:1}. 
Finally we note that 
$$\P \textrm{-a.s.} \quad \forall \no\in \No \quad \mathcal P(\Xc^\no)=\E_\P[\Xc^\no|\G_{-1}]=\E_{\P_{\omega}}[\Xc^\no]$$
and therefore there exists a set $\Omega_\Xc\in\G_{-1}$ such that $\P(\Omega_\Xc)=1$ and, for every $\omega\in \Omega_\Xc$ it holds that 
$\mathcal P(\Xc^\no)=\E_{\P_{\omega}}[\Xc^\no]$ for each $\no\in \No$.
To complete that proof it is enough to take $\Omega^\P=\Omega_{-1}\cap\Omega_m\cap\Omega_\Xc$.
\finproof

{\sc Proof of Lemma \ref{classical}:}\\
Using Propositions \ref{hedge} and \ref{price}, it is enough to show the asserted inequality separately on each atom $A^\omega$ of the $\sigma$-field $\G_{-1}$. 
The proof then follows by a  classical argument. 
Take any $\gg$-admissible super--replicating portfolio $(\mo, \gamma)\in \mathcal A_{\Xc}(\gg)$ on $A^\omega$ and any measure $\P\in \mathcal M^{\gg}_{\mathcal X, \mathcal P, A^\omega}$. 
Let $\{\P_{v}\}$ denote regular conditional probabilities of $\P$ with respect to $\G_{0}$.
Thus, by Lemma \ref{rcp}, $\P$-a.s., $\P_v\in \mathcal M^{\gg}_{B^v}$ where $\{B^v\}$ are $\G_0$-atoms containing $v$. 
Note that $\{B^v\}$ form finer partition than $\{A^\omega\}$.
Then, since $\P$-a.s., $\P_v(M\equiv const)=1$ and $\P_v(\gamma=\gamma\id_{B^v})$ where $\gamma\id_{B^v}$ is jointly measurable, we deduce that 
$$\E_{\P_v}[\xi]\leq \E_{\P_v}\left[\mo\Xc+\int_0^T\gamma_u dS_u\right]\leq \E_{\P_v}[\mo\Xc]\quad \P\textrm{-a.s.}$$
as $\int_0^\cdot\gamma_u\id_{B^v}d S_u$ is a $\gg$-local martingale bounded from below and thus a $\gg$-super\-mar\-tin\-ga\-le. 
Since $\{\P_v\}$ are regular conditional probabilities of $\P$ and $\P(\mo^\no=\mo^\no(\omega))$ for each $\no\in\No$, taking expectations under $\P$, we have
$$\E_{\P}[\xi]\leq 
\E_{\P}[\mo\Xc]=\mo(\omega)\mathcal P(\Xc)$$
which completes the proof.
\finproof

{\sc Proof of Lemma \ref{alfa_m}:}\\
Note that $\ale^v$ can be written as
\begin{align*}
\ale^{v}(\wt v, \omega):=
\begin{cases}
\ale_1(\wt v,\omega) & \textrm{on}\, \, \Omega_{+}\times\Omega_{+}\backslash \{(\wt v, \omega): \omega\vert_{[0,T_1]}=\wt v\vert_{[0,T_1]}\;\lor\; \omega\vert_{[0,T_1]}=v\vert_{[0,T_1]}\}\\
\ale_2(\omega) &\textrm{on}\, \,  \{(\wt v, \omega): \omega\vert_{[0,T_1]}=\wt v\vert_{[0,T_1]}\}\\
\ale_3(\wt v) & \textrm{on}\, \,  \{(\wt v, \omega): \omega\vert_{[0,T_1]}=v\vert_{[0,T_1]}\}
\end{cases}
\end{align*}
where $\ale_i$ for $i\in\{1,2,3\}$ 
are given by $\ale_1(\wt v, \omega)=\omega$, $\ale_2(\omega)=\ale^v(\omega,\omega)=v\vert_{[0, T_1]}\otimes \frac{v_{T_1}}{\omega_{T_1}}  \omega\vert_{[T_1, T]}$ and $\ale_3(\omega)=\ale^v(\omega,v)=\omega\vert_{[0, T_1]}\otimes \frac{\omega_{T_1}}{v_{T_1}} \;v\vert_{[T_1, T]} $.
Thus, since sets $\{(\wt v, \omega): \omega\vert_{[0,T_1]}=\wt v\vert_{[0,T_1]}\}$ and $\{(\wt v, \omega): \omega\vert_{[0,T_1]}=v\vert_{[0,T_1]}\}$ are $\F_T\otimes\F_T$-measurable, it is now enough to show that each $\ale_i$ is measurable. 
The map $\ale_1$ is measurable since it is simply a projection.
The mappings $\ale_2, \ale_3: \Omega_{+}\to \Omega_{+}$ are continuous thus measurable. 
Indeed, fix $\omega\in\Omega_{+}$ and consider $\wt \omega$ such that $\norm{\wt \omega-\omega}\leq \delta\leq \frac{1}{2}\omega_{T_1}$. Then
\begin{align*}
\norm{\ale_2(\omega)-\ale_2(\wt \omega)}
&=\sup_{t\in[T_1,T]}\left|\frac{v_{T_1}}{\wt \omega_{T_1}}(\omega_t-\wt \omega_t)+\frac{v_{T_1}}{\wt \omega_{T_1}\omega_{T_1}}(\omega_{T_1}-\wt \omega_{T_1})\omega_t\right|
\leq \frac{2v_{T_1}}{\omega_{T_1}}\left (1+\frac{\norm{\omega}}{\omega_{T_1}}\right)\delta,\\
\norm{\ale_3(\omega)-\ale_3(\wt \omega)}
&=\norm{\omega-\wt \omega}\lor\sup_{t\in[T_1,T]}\left|\frac{v_{t}}{v_{T_1}}(\omega_{T_1}-\wt \omega_{T_1})\right|
\leq \frac{\norm{v}}{v_{T_1}}\delta.
\end{align*}
\finproof

{\sc Proof of Proposition \ref{dpV}:}\\
In the proof we denote $\wt \xi:=V^{\gg, [T_1, T]}_\Omega(\xi)$. \\
(i) Let $v:=(v^1,...,v^d)\in \Omega$ and $\wt v:=(\wt v^1,..., \wt v^d) \in \Omega$. 
Note that 
\begin{align*}
\left |\wt \xi((v^{1},.., v^{d}))-\wt \xi((\wt v^{1},.., \wt v^{d}))\right | 
\leq \sum_{k=1}^d \left |\wt \xi((\wt v^{1},..,\wt v^{k-1}, v^k, .., v^{d}))-\wt \xi((\wt v^{1},.., \wt v^{k}, v^{k+1},.., v^{d}))\right |.
\end{align*}
Thus, to establish uniform continuity of $\wt \xi$, it is enough to consider $v$ and $\wt v$ which differ on one coordinate only and, without loss of generality, we may assume that $d=1$.

Consider a small $\delta>0$. Suppose that $\norm{v-\wt v}_{[0,T_1]}\leq \delta$, $|v_{T_1}- \wt v_{T_1}|=\d\ge 0$, $v_{T_1}>0$ and $\wt v_{T_1}>0$.\\
In the first step we show that $\wt \xi(\wt v)\leq \wt \xi(v)+\varepsilon$ for an appropriately chosen $\varepsilon$, depending only on $\xi$ and $\delta$.
For each $\eta>0$ there exists a strategy $\gamma$ such that 
$$\wt \xi(v)+\int_{T_1}^T\gamma_t d S_t\geq \xi-\eta\textrm{ on } B^{v}.$$ 
Let $\lambda = v_{T_1}/\wt v_{T_1}\in (0, \infty)$ and
define the path modification mapping $\ale$ by $\ale(\omega):=\ale^{v}(\wt v,\omega)$ where $\ale^v$ is given in \eqref{alfa}.
Note that $\ale$ is a bijection satisfying  
$\ale=\ale^{-1}$. 
Introduce a stopping time
\begin{align}
\label{wttau}
\wt \tau(\omega) := \tau^{v, \wt v}(\omega):=
\begin{cases}
\inf\{t> T_1:\, \omega_t - \wt v_{T_1} \ge \wt v_{T_1}\d^{-\frac{1}{2}}\}\land T
& \omega \in B^{\wt v}\\
\inf\{t> T_1:\, \omega_t - v_{T_1} \ge v_{T_1}\d^{-\frac{1}{2}}\}\land T & \omega \in B^{v}\\
T_1 & \omega \notin B^{v}\cup B^{\wt v}
\end{cases}
.
\end{align}
To show that $\wt \xi(\wt v)\leq \wt \xi(v)+\varepsilon$ we will consider a strategy $\lambda \gamma\circ \ale+\frac{\d^{\frac{1}{4}}}{\wt v_{T_1}}\indicators{[T_1, {\wt \tau})}$ on $B^{\wt v}$. 
The second term of this strategy is clearly $\gg$-adapted. To show that the first term is $\gg$-adapted as well, it is enough to show that $Z\circ \ale$ is $\sigma(Z)$-measurable. The last is true since 
$$Z\circ\ale(\omega)=\wt Z\left (\frac{\ale(\omega)\vert_{[T_1, T]}}{\ale(\omega)_{T_1}}\right)
=\wt Z\left (\frac{\omega\vert_{[T_1, T]}}{\omega_{T_1}}\right)=Z(\omega).$$
Then, we obtain
\begin{align*}
\wt \xi(v)&+\lambda\int_{T_1}^T\gamma\circ \ale (\omega)_td S_t(\omega)+\frac{\d^{\frac{1}{4}}}{\wt v_{T_1}}(\omega_{\wt \tau}-\wt v_{T_1})\\
&=\wt \xi(v)+\int_{T_1}^T\gamma\circ \ale (\omega)_td S_t\circ\ale (\omega)+\frac{\d^{\frac{1}{4}}}{\wt v_{T_1}}(\omega_{\wt \tau}-\wt v_{T_1})\\
&\geq \xi\circ\ale(\omega)-\eta
+\frac{\d^{\frac{1}{4}}}{\wt v_{T_1}}(\omega_{\wt \tau}-\wt v_{T_1})
\end{align*}
where the first equality is due to our definition of integration. 
In the case that ${\wt \tau}(\omega)=T$ one has 
$$\frac{\d^{\frac{1}{4}}}{\wt v_{T_1}}(\omega_{\wt \tau} - \wt v_{T_1})\ge -\frac{\d^{\frac{1}{4}}}{\wt v_{T_1}}\wt v_{T_1} = -\d^{1/4}$$
and
and
\begin{equation}
\label{omegabeta}
\|\ale(\omega) - \omega\|\le \delta \lor (\lambda-1)\wt v_{T_1}(\d^{-\frac{1}{2}}+1)\le 2\delta^{1/2}.
\end{equation}
Thus, for $\wt \tau(\omega)=T$, it follows that
\begin{align*}
\xi\circ\ale(\omega)-\eta
+\frac{\d^{\frac{1}{4}}}{\wt v_{T_1}}(\omega_\tau-\wt v_{T_1})
\ge \, \xi(\omega) - e_{\xi}(2\delta^{1/2})-\eta
-\d^{1/4} 
\end{align*}
where $e_\xi$ is modulus of continuity of $\xi$.
Hence, for ${\wt \tau}(\omega)=T$, we deduce $\wt \xi(\wt v)\leq \wt \xi(v)+ e_{\xi}(2\delta^{1/2})+\d^{1/4}$.
In the case that ${\wt \tau}(\omega)<T$ one has 
$$ \frac{\d^{\frac{1}{4}}}{\wt v_{T_1}}(\omega_{\wt \tau} - \wt v_{T_1}) = \frac{\d^{\frac{1}{4}}}{\wt v_{T_1}}\wt v_{T_1}\d^{-\frac{1}{2}} = \d^{-1/4}$$
and 
\begin{align*}
\xi\circ\ale(\omega)-\eta
+\frac{\d^{\frac{1}{4}}}{\wt v_{T_1}}(\omega_\tau-\wt v_{T_1})
\ge\, -\norm{\xi}-\eta +\d^{-1/4}
\end{align*}
which, for $\d$ small enough ($\d\leq (2\norm{\xi})^{-4}$), dominates $\xi(\omega)$. We deduce that $\wt \xi(\wt v)\leq \wt \xi(v)+ e_{\xi}(2\delta^{1/2})+\d^{1/4}$ and conclude that $\wt \xi$ is uniformly continuous on $\{\omega\in \Omega\,:\, \|\omega\|> 0\}$.\\
To complete the proof, we now consider the case where $\wt v_{T_1}=0$. 
Let, for some small $\delta>0$, $\norm{v-\wt v}_{[0,T_1]}\leq \delta$ and $v_{T_1}=\d>0$. 
Firstly notice that $\wt v$ must satisfy $\wt \xi(\wt v) = \xi(\wt v\vert_{[0, T_1]}\otimes 0\vert_{[T_1, T]})$ since we can buy any amount of stock at price 0 at time $T_1$ thus only constant path is relevant, and therefore $\wt \xi(\wt v)\leq \wt \xi(v)+ e_{\xi}(\delta)$. Now consider the strategy $\gamma$ for $\omega\in B^{v}$ defined as
$$\gamma(\omega) := \delta^{-1/2}\indicators{[T_1, \sigma(\omega))} 
\;\;\textrm{ where } \;\;\sigma(\omega) := \inf\{t> T_1:\, \omega_t - v_{T_1} \ge \delta^{1/4}\}.$$ 
Then, whenever $\sigma(\omega)< T$, 
$$ \wt \xi(\wt v)+\int_{T_1}^T\gamma(\omega)_tdS_t(\omega) = \wt \xi(\wt v)+\delta^{-1/2}(\omega_{\sigma} - v_{T_1})
=\wt \xi(\wt v)+\delta^{-1/4}$$
which, for $d$ small enough, majorates $\xi(\omega)$.
Otherwise, if $\sigma(\omega)=T$ 
$$\wt \xi(\wt v)+\int_{T_1}^T\gamma(\omega)_tdS_t(\omega) \ge  \wt \xi(\wt v)-\delta^{1/2}\geq \xi(\omega)- e_{\xi}(2\delta^{1/4})-\delta^{1/2}$$
since $\norm{\wt v-\omega}\leq 2\delta^{1/4}$. 
Therefore, $\wt \xi(v)\leq  \wt \xi(\wt v)+ e_{\xi}(2\delta^{1/4}) + \delta^{1/2}$.\\

(ii) Let $V^1:=V^{\gg, [0,T]}_\I(\xi)$ and $V^2:=V^{\gg, [0,T_1]}_\I(\wt \xi)$.
For each $\eta>0$ there exists a strategy $\gamma\in \mathcal A(\gg, [0,T])$ such that 
$$V^1+\int_{0}^{T_1}\gamma_t d S_t+\int_{T_1}^{T}\gamma_t d S_t\geq \xi -\eta\quad \textrm{on} \quad \Omega.$$ 
Let $\tau(S) := \inf\{t > 0\,:\, V^1 + \int_0^{t}\gamma_u d S_u \ge \sup_{\omega\in \Omega}\xi(\omega) -\eta\}\wedge T$. It is a stopping time. Hence $\wt{\gamma}:=\gamma\indicators{[0,\tau]}\in \mathcal A(\gg, [0,T])$ and satisfies that
 $$V^1+\int_{0}^{T_1}\wt{\gamma}_t d S_t+\int_{T_1}^{T}\wt{\gamma}_t d S_t\geq \xi -\eta\quad \textrm{on} \quad \Omega.$$ 
Moreover, for any $t\ge T_1$, 
$$\int_{T_1}^{t}\wt{\gamma}_u d S_u\geq \xi - \sup_{\omega\in \Omega}\xi(\omega) \quad \textrm{on} \quad \Omega,$$
and therefore $\wt{\gamma}\in \mathcal A(\gg, [T_1,T])$. 
 
In particular, for a fixed $\omega\in \Omega$, the superhedging holds on $B^v$.
Since $V^1+\int_{0}^{T_1}\wt{\gamma}_t d S_t$ is constant on $B^\omega$, we deduce that 
$V^1+\int_{0}^{T_1}\wt{\gamma}_t d S_t\geq \wt \xi$ on $\Omega\vert_{[0, T_1]}$ and therefore $V^1\geq V^2$.\\

To prove the reverse inequality take  $z> V^2$. First, there exists $\gamma^{1}\in \mathcal A(\ff, [0,T_1])$ such that $z+\int_{0}^{T_1}\gamma^1_t d S_t\geq \wt \xi$ on $\Omega$. 
If, for each $\eta>0$, there exists a strategy $\gamma^{2}\in \mathcal A(\gg, [T_1, T])$ such that $\gamma^2$ is jointly measurable and $z+\int_{0}^{T_1}\gamma^{1}_t d S_t+\int_{T_1}^{T}\gamma^{2}_t d S_t\geq \xi-\eta$, then clearly $z\geq V^1$. 

We now show the existence of such $\gamma^{2}$ for every $\eta>0$. Let $\{\omega^n\}_n$ be a countable dense subset of $\Omega |_{[0, T_1]}$ and $B^n:=B^{\omega^n}$, and denote the closed ball around $\omega^n$ of radius $\delta$ by $\wt B^n(\delta):=\{\omega: \sup_{t\in [0, T_1]}|\omega_t-\omega^n_t|\leq \delta\}$. 
Define the path modification mapping $\ale^{n, \omega}$ by $\ale^{n, \omega}:=\ale^{\omega^n}( \omega,\cdot)$ where $\ale^{\omega^n}( \omega,\cdot)$ is given in \eqref{alfa}.
Note that $\ale^{n, \omega}$ is a bijection satisfying  
$\ale^{n, \omega}=(\ale^{n, \omega})^{-1}$. 
We now take $\{\gamma^n\}_n$, a set of strategies in $\mathcal A(\gg, [T_1, T])$, such that 
\begin{equation}
\wt \xi(\omega^n) + \int_{T_1}^T\gamma^n_u(S) d S_u \ge \xi - \delta \quad \text{ on $B^n$. } \label{eq: admissibility_each}
\end{equation}
Let us consider $\wt{\gamma}^n:\Omega\to $ defined by $\wt{\gamma}_n(\omega) = 0 $ if $\omega\not\in \wt B^n(D)$ and for $\omega\in \wt B^n(D)$
\begin{align*}
\wt{\gamma}_n(\omega):=
\begin{cases}
\frac{\omega^n_{T_1}}{\omega_{T_1}} \gamma^n\circ \ale^{n,\omega}+\frac{\d^{\frac{1}{4}}}{\omega_{T_1}}\indicators{[T_1, \tau^{\omega^n,\omega})} &\mbox{ if  $\omega^n_{T_1} \ge \delta$, } \\
\delta^{-1/2}\indicators{[T_1, \sigma(\omega))} &\mbox{ if  $\omega^n_{T_1} < \delta$,}
\end{cases}
\end{align*} 
where $\sigma(\omega) := \inf\{t> T_1:\, \omega_t - \omega_{T_1} \ge \delta^{1/4}\}$.
It follows from above that there exists a constant $\epsilon(D, \delta)$ which depends on $D$ and $\delta$ with $\epsilon(D, \delta) \to 0$ as $D, \delta\to 0$, such that 
$$\wt{\xi}(S) + \int_{T_1}^{T}\wt{\gamma}^n_u(S) d S_u \ge \xi(S) - \epsilon(D, \delta).$$
The strategy $\wt{\gamma}^n$ is clearly $\F_{T}$-measurable. We also notice that it is adapted to $\F$ on $[T_1, T]$ since it is straightforward to see that for any $\omega,\upsilon\in \Omega$ such that $\omega_u = \upsilon_u$ for any $u\le [t, T]$ with $t\ge T_1$, $\wt{\gamma}^n_u(\omega) = \wt{\gamma}_u(\upsilon)$ on $[T_1, t]$. Hence, $\wt{\gamma}^n\in \AA([T_1, T])$. In addition, we know that for any $n$, $t\in [T_1, T]$ and $S\in B^n$, there exists $\wt{S}$ such that $\wt{S}_u = S_u$ for any $u\le t$ and $\wt{S}_u = S_t$ for any $u\ge t$, and therefore
\begin{align}
\int_{T_1}^t\gamma^n_u(S) d S_u  = \int_{T_1}^T\gamma^n_u(\tilde{S}) d \tilde{S}_u\ge \xi(\wt{S}) - \delta - \wt \xi(\omega^n)\ge 2\inf_{\omega\in \Omega}{\xi(\omega)} - 1.
\end{align}

Let us now define $\wt{\gamma}^\varepsilon$ by
\begin{equation*}
\wt{\gamma}^\varepsilon(\omega):=\sum_{n}\id_{C^n}(\omega)\wt{\gamma}^n(\omega)\quad \textrm{where}\quad {C^n:=\wt B^n\backslash \bigcup_{k=1}^{n-1}\wt B^k}.
\end{equation*}

It is then straightforward to see that $\wt{\gamma}^\varepsilon$ is progressively measurable and satisfies the admissibility condition in \eqref{eq:admissibility}. 

\finproof

{\sc Proof of Proposition ref{dpP}:}\\
In the proof we denote $\wh \xi:=P^{\gg, [T_1, T]}_\Omega(\xi)$.\\
(i) Let $v:=(v^1,...,v^d)\in \Omega$ and $\wt v:=(\wt v^1,..., \wt v^d) \in \Omega$. 
Note that 
\begin{align*}
\left |\wh \xi((v^{1},.., v^{d}))-\wh \xi((\wt v^{1},.., \wt v^{d}))\right | 
\leq \sum_{k=1}^d \left |\wh \xi((\wt v^{1},..,\wt v^{k-1}, v^k, .., v^{d}))-\wh \xi((\wt v^{1},.., \wt v^{k}, v^{k+1},.., v^{d}))\right |.
\end{align*}
Thus, to prove uniform continuity of $\wt \xi$, it is enough to consider $v$ and $\wt v$ which differ on one coordinate only and, without loss of generality, we may assume that $d=1$. \\
Suppose that $\norm{v-\wt v}_{[0,T_1]}\leq \delta$ and $|v_{T_1}- \wt v_{T_1}|=\d\geq 0$.
It is enough to show that $\wh \xi(\wt v)\leq \wh \xi(v)+\varepsilon$ for an appropriately chosen $\varepsilon$ depending only on $\xi$ and $\delta$.
Take $\P\in \mathcal M^{\gg, [T_1, T]}_{B^{\wt v}}$, i.e.: $\P(B^{\wt v})=1$, where $B^{\wt v}:=\{\omega: \omega_t=\wt v_t \textrm{  for  }t\in [0, T_1]\}$, $\P(S_{T_1}=\wt v_{T_1})=1$ and $\E_\P[S_t\id_G]=\E_\P[S_s\id_G]$ for each $T_1\leq s\leq t\leq T$ and $G\in \G_s$. 
Define measure $\bar \P$ as $\bar \P=\P\circ \ale$ with path modification $\ale$ given in \eqref{alfa}.
Then $\bar \P$ is an element of $\mathcal M^{\gg, [T_1, T]}_{B^{v}}$ since 
$\bar \P(B^{v})=\P(\ale(B^{v}))= \P(B^{\wt v})=1$, it is a martingale measure on $[T_1, T]$ as for $T_1\leq s\leq t\leq T$ and $G\in \G_s$
\begin{align*}
\E_{\bar \P}[S_t\id_{G}]
&=\E_{\P}[(S\circ \ale)_t\id_{\ale(G)}]
=\E_{\P}[(S\circ \ale)_s\id_{\ale(G)}]
=\E_{\bar \P}[S_s\id_{G}],
\end{align*}
where the second equality follow by $\ale(G)\in \G_s$. The latter is true since, for any Borel set $B$,
one has $\ale(\{Z\in B\}\cap B^v)=B^{\wt v}\cap \{Z\in A\}$; the $\sigma$-field $\F_s$ coincides with trivial $\sigma$-field up to $\P$-null sets and up to $\bar \P$-null sets; the general case follows from the monotone class argument.
Hence, with $\wt \tau$ defined in \eqref{wttau},
\begin{align}
\nonumber
|\E_{\P}[\xi] - \E_{\bar \P}[\xi]| 
&=|\E_{\P}[\xi] - \E_{\P}[\xi\circ \ale]| \\
\nonumber 
&=\E_{\P}[(\xi - \xi\circ \ale)\indicators{\{\wt\tau=T\}}] +\E_{\P}[(\xi - \xi\circ \ale)\indicators{\{\wt\tau<T\}}] \\
\label{blisko}
&\le e_{\xi}(2\delta^{1/2}) + 2\norm{\xi} \frac{\d}{\d+\d^{1/2}},
\end{align}
where in the last inequality we used \eqref{omegabeta}, Doob's inequality and the fact that
\[
\P(\wt\tau<T)
=\P\left(\sup_{t\in[T_1, T]}S_t\geq \wt v_{T_1}(1+\d^{-1/2})\right)
\leq \frac{\wt v_{T_1}}{\wt v_{T_1}(1+\d^{-1/2})}=\frac{\d}{\d+\d^{1/2}}.
\] 
(ii) To prove that $P^{\gg, [0,T]}_\I(\xi)\leq P^{\ff, [0,T_1]}_\I(P^{\gg, [T_1, T]}_\Omega(\xi))$ it is enough to note that: 
$$\sup_{\P\in \mathcal M^{\gg, [0, T]}_\Omega} \E_\P[\xi]
=\sup_{\P\in \mathcal M^{\gg, [0, T]}_\Omega} \E_\P[\E_{\P_\omega}[\xi]]
\leq \sup_{\P\in \mathcal M^{\gg, [0, T]}_\Omega} \E_\P\left [\sup_{\bar \P\in \mathcal M^{\gg, [T_1, T]}_{B^\omega}}\E_{\bar \P}[\xi]\right ]$$ where $\{\P_\omega\}$ is regular conditional distribution with respect to $\F_{T_1}$ and where in the last step we used measurability implied by assertion (i).\\
Now we will show the remaining inequality. 
Let $\{\omega^n\}_n$ be a countable dense subset of $\Omega |_{[0, T_1]}$ and $B^n:=B^{\omega^n}$.
Define the path modification mapping $\ale^{n, \omega}$ by $\ale^{n, \omega}:=\ale^{\omega^n}(\omega,\cdot)$ where $\ale^{\omega^n}(\omega,\cdot)$ is given in \eqref{alfa}.
Note that $\ale^{n, \omega}$ is a bijection satisfying  
$\ale^{n, \omega}=(\ale^{n, \omega})^{-1}$. 
For any $\P_n\in \mathcal M^{\gg, [T_1, T]}_{B^{n}}$ the measure $\P_n\circ \ale^{n, \omega}$ belongs to $\mathcal M^{\gg, [T_1, T]}_{B^{\omega}}$. 
Moreover, similarly to \eqref{blisko}, we obtain that
\[
|\E_{\P_n}[\xi] - \E_{\P_n\circ \ale^{n, \omega}}[\xi]| \le e_{\xi}(2\delta^{1/2}) + 2\norm{\xi} \frac{\delta}{\delta+\delta^{1/2}}
\]
whenever $\norm{\omega^n-\omega}_{[0,T_1]}\leq \delta$.
Let us consider probability kernel $N_n:\Omega\to \mathcal M^{\gg, [T_1, T]}$ defined by $N_n(\omega):=\P_n\circ \ale^{n, \omega}$. 
The kernel $N_n$ is $\F_{T}$-measurable, i.e., $N_n(\omega, F)=\P_n\circ \ale^{n, \omega}(F)$ is $\F_{T}$-measurable for any $F\in \F_T$, since $(\omega,\wt \omega)\to \id_F\circ \ale^{n,\omega}(\wt \omega)$ is $\F_T\otimes\F_T$-measurable and bounded thus $\E_{\P_n}[\id_F\circ \ale^{n,\omega}]$ is $\F_T$-measurable (see \cite[Section 3.3]{bogachev}). Measurability of $\ale^{n}(\wt \omega, \omega)$ was shown in Lemma \ref{alfa_m}.
Then, since $N_n$ is constant on atoms of $\F_{T_1}$, we deduce from Blackwell's Theorem (see \cite[Theorem 8.6.7]{cohn} and/or \cite[Ch III, \textsection 26, p.80-81]{dell14}) that $N_n$ is $\F_{T_1}$-measurable probability kernel.

\noindent Denoting the closed ball around $\omega^n$ of radius $\delta$ by $\wt B^n(\delta):=\{\omega: \sup_{t\in [0, T_1]}|\omega_t-\omega^n_t|\leq \delta\}$, we observe that
\[
\sup_{\P\in \mathcal M^{\gg, [T_1,T]}_{\wt B^n(\delta)}}\E_\P[\xi]
=\sup_{\omega\in \wt B^n(\delta)}\sup_{\P\in \mathcal M^{\gg, [T_1,T]}_{B^\omega}}\E_\P[\xi]
=\sup_{\omega\in \wt B^n(\delta)}\wh \xi(\omega)\leq \wh \xi(\omega^n)+\varepsilon(\delta)
\]
where the second equality follows from uniform continuity of $\wh \xi$. 

Fix $\varepsilon>0$. Then we can chose $\delta>0$ and family of measures $\P_n^\varepsilon\in \mathcal M^{\gg, [T_1, T]}_{B^{n}}$ for each $n$ such that 
$$\varepsilon/2 +\E_{\P_n^\varepsilon}[\xi]\geq\wh \xi(\omega)\quad \forall \omega\in \wt B^n(\delta)\quad \textrm{and}\quad e_{\xi}(2\delta^{1/2}) + 2\norm{\xi} {\delta}/(\delta+\delta^{1/2})\leq \varepsilon/2.$$
Let us now define the $\F_{T_1}$-measurable probability kernel $N^\varepsilon$ as
\[
N^\varepsilon(\omega):=\sum_n \id_{C^n}(\omega)\;\P^\varepsilon_n\circ \ale^{n, \omega}\quad \textrm{where}\quad {C^n:=\wt B^n\backslash \bigcup_{k=1}^{n-1}\wt B^k}.
\]
Probability kernel $N^\varepsilon$ is constructed such that it satisfies 
\[\varepsilon +\E_{N^\varepsilon(\omega)} [\xi]\geq \wh \xi(\omega)\quad \forall \omega\in \Omega\quad \textrm{and}\quad N^\varepsilon(\omega)\in \mathcal M^{\gg, [0,T_1]}_{B^\omega}.
\]
There as well exists a measure $\P^\varepsilon\in \mathcal M^{\ff, [0, T_1]}_\Omega$ such that $\varepsilon +\E_{\P^\varepsilon}[\wh\xi]\geq \sup_{\P\in \mathcal M^{\ff, [0,T_1]}_{\Omega}}\E_{\P}[\wh\xi]$.
The concatenation of measures $\bar \P^\varepsilon:= \P^\varepsilon\otimes N^\varepsilon$ (see Section 3.1 in \cite{kt:cap}), defined, for each $F\in \F_T$, as 
$$\bar \P^\varepsilon(F)=\E_{\P^\varepsilon}\Big[\sum_n \id_{{C^n}}N^\varepsilon(F)\Big]$$
is a probability measure.
Note that regular conditional probabilities of $\bar \P^\varepsilon$ w.r.t $\F_{T_1}$ equal to $N^\varepsilon$ 
 and $d\,\bar\P^\varepsilon\vert_{\F_{T_1}}=d\,\P^\varepsilon\vert_{\F_{T_1}}$. Thus, for $s\leq t$ and $G_s\in \G_s$, we have
\begin{align*}
\E_{\bar \P^\varepsilon}\big[(S_t-S_s)\id_{G_s}\big]
&=\E_{\bar \P^\varepsilon}\big[\E_{N^\varepsilon}[(S_t-S_s)\id_{G_s}]\big]\\
&=\E_{\bar \P^\varepsilon}\big[\E_{N^\varepsilon}[(S_{s\lor T_1}-S_s)\id_{G_s}]\big]\\
&=\E_{\bar \P^\varepsilon}\big[(S_{s\lor T_1}-S_s)\id_{G_s}\big]\\
&=\E_{\P^\varepsilon}\big[(S_{s\lor T_1}-S_s)\id_{G_s}\big]\\
&=0
\end{align*}
which shows that $S$ is a $(\bar \P^\varepsilon, \gg)$-martingale. 
Moreover $\bar \P^\varepsilon$ satisfies
$$\E_{\bar \P^\varepsilon}[\xi]
=\E_{\P^\varepsilon}\big[\E_{N^\varepsilon}[\xi]\big]
\geq \E_{ \P^\varepsilon}[\wh \xi]-\varepsilon
\geq \sup_{\P\in \mathcal M^{\ff, [0, T_1]}_{\Omega}}\E_{\P}[\wh\xi]-2\varepsilon.$$
The proof is completed.
\finproof

\bibliographystyle{plaintest}
\bibliography{biblio_robust}

\end{document}